\documentclass{osa-article}
\journal{osajournal}
\articletype{Research Article}

\begin{document}

\title{Standardized spectral and radiometric calibration of consumer cameras}

\author{Olivier Burggraaff\authormark{1,2,*}, Norbert Schmidt\authormark{3}, Jaime Zamorano\authormark{4}, Klaas Pauly\authormark{5}, Sergio Pascual\authormark{4}, Carlos Tapia\authormark{4}, Evangelos Spyrakos\authormark{6}, and Frans Snik\authormark{1}}

\address{\authormark{1}Leiden Observatory, Leiden University, PO Box 9513, 2300 RA Leiden, The Netherlands\\
\authormark{2}Institute of Environmental Sciences (CML), Leiden University, PO Box 9518, 2300 RA Leiden, The Netherlands\\
\authormark{3}DDQ Apps, Webservices, Project Management, Maastricht, The Netherlands\\
\authormark{4}Dept. Astrof\'isica y CC. de la Atm\'osfera, IPARCOS, Universidad Complutense de Madrid, Madrid, 28040, Spain\\
\authormark{5}VITO; Flemish Institute for Technological Research, Mol, Belgium\\
\authormark{6}Biological and Environmental Sciences, School of Natural Sciences, University of Stirling, Stirling, United Kingdom}

\email{\authormark{*}burggraaff@strw.leidenuniv.nl}

\begin{abstract}
Consumer cameras, particularly onboard smartphones and UAVs, are now commonly used as scientific instruments. However, their data processing pipelines are not optimized for quantitative radiometry and their calibration is more complex than that of scientific cameras. The lack of a standardized calibration methodology limits the interoperability between devices and, in the ever-changing market, ultimately the lifespan of projects using them. We present a standardized methodology and database (SPECTACLE) for spectral and radiometric calibrations of consumer cameras, including linearity, bias variations, read-out noise, dark current, ISO speed and gain, flat-field, and RGB spectral response. This includes golden standard ground-truth methods and do-it-yourself methods suitable for non-experts. Applying this methodology to seven popular cameras, we found high linearity in RAW but not JPEG data, inter-pixel gain variations >400\% correlated with large-scale bias and read-out noise patterns, non-trivial ISO speed normalization functions, flat-field correction factors varying by up to 2.79 over the field of view, and both similarities and differences in spectral response. Moreover, these results differed wildly between camera models, highlighting the importance of standardization and a centralized database.
\end{abstract}

\section{Introduction} \label{s:intro}

Consumer cameras have seen increasing scientific use in recent years. Their low cost makes them ideal for projects involving large scale deployment, autonomous monitoring, or citizen science. Successful scientific applications include environmental monitoring \cite{SanchezdeMiguel2019artificiallight, Leeuw2018hydrocolor, Yang2018HydroColor, Gallagher2018hydrocolor, Igoe2018ozone, Haenel2018DSM, Cai2017addon, Friedrichs2017SmartFluo, Novoa2015WACODI, Snik2014ispex, Sumriddetchkajorn2014chlorine, Kreuter2010skypolarization, Goddijn2009fundamentals}, cosmic ray detection \cite{Whiteson2016cosmicrays}, vegetation mapping \cite{Ruwaimana2018droneadvantages, Rasmussen2016UAVvegetation, Flynn2014UAVaquaticvegetation, Liang2014treemapping, Lebourgeois2008commercialcameras}, color science \cite{Kordecki2017vignettingSPIE, Novoa2015WACODI, Nguyen2014raw2raw, Charriere2013rgbmicroscope, Jiang2013spectralsensitivity, Bongiorno2013spectralresponse, Cheung2005multispectralcharacterization}, and biomedical applications \cite{Khedekar2019eyecancer, Mannino2018anemia, Ding2018mHealth, Wang2017biospectrometer, Long2017TRI, Lam2015heartrate, Skandarajah2014microscope, Smith2011microspectro}. However, the use of consumer cameras is made difficult by limited software controls and camera specifications. Inter-calibration of multiple camera models is complex and laborious and the market constantly shifting, and for these reasons many projects are limited to only a few devices. These constraints severely affect both the data quality and the sustainability of projects using consumer cameras.

Smartphones, in particular, have become a common tool for research, thanks to their wide availability and features such as wireless connectivity. Many scientific applications (apps) using smartphone cameras have been developed, across a variety of fields. A recent example is HydroColor, a citizen science tool for measuring water quality, specifically turbidity and remote sensing reflectance $R_{rs}$. These are derived from RGB color photographs using standard inversion algorithms. Results from this app agree well with professional standard equipment, with mean errors in $R_{rs}$ and turbidity $\leq$26\% compared to reference sensors. However, due to software constraints, the app uses compressed JPEG data rather than raw sensor data and assumes identical spectral responses for all cameras. These factors severely limit the possible data quality. Nevertheless, HydroColor has already seen significant adoption by the community, and future developments may reduce the aforementioned limitations \cite{Leeuw2018hydrocolor, Yang2018HydroColor, Gallagher2018hydrocolor}. Another recent application of smartphone cameras is bioluminescent-based analyte quantitation by smartphone (BAQS), a technique for the detection of bioluminescent bacteria. Using BAQS, flux intensities down to the pW scale can be detected on some smartphone models; however, on others, software constraints and dark noise severely limit its sensitivity \cite{Kim2017smartphonebioluminescence}. As a final example, Skandarajah et al. used smartphone cameras with conventional microscopes for micron-scale imaging, for example of stained blood samples. Resolutions comparable to scientific cameras were achieved, but intensity and color measurements were limited by a lack of camera control and factors including nonlinearity and white balance \cite{Skandarajah2014microscope}. A full review of smartphone science is outside the scope of this work, and we instead refer the reader to a number of extensive reviews by other authors \cite{Grossi2019smartphonesensors, Zaidan2018smartphoneskincancer, Kanchi2018smartphonebioreview, Huang2018smartphonebiosensors, McGonigle2018spectrometers, Rateni2017smartphonefoodreview, Li2016SmartphonesSensing, McCracken2016resourcelimited}. 

Smartphone spectroscopy is an active field of development \cite{McGonigle2018spectrometers, Crocombe2018spectroscopy}. Many spectroscopic add-ons have been developed, including do-it-yourself models costing less than \$10 at Public Lab (\url{https://publiclab.org/wiki/spectrometry}). One early smartphone spectrometer was iSPEX, a spectropolarimetric add-on for iPhone devices used by >3000 citizen scientists to measure aerosol optical thickness (AOT) in the Netherlands in 2013. iSPEX data were found to agree well with reference sensors, with a correlation coefficient of $r = 0.81$ between AOT values observed with iSPEX and with the Moderate Resolution Imaging Spectroradiometer (MODIS) Aqua and Terra satellites \cite{Snik2014ispex}. However, the iSPEX data were limited in their polarimetric accuracy (absolute uncertainties in the degree of linear polarization (DoLP) $\approx 0.03$), preventing quantitative measurements of aerosol compositions and sizes \cite{Snik2014ispex}. This relatively large error stemmed from a lack of camera controls, such as the inability to fix the focus of the camera to a controlled and reproducible position. Furthermore, the sustainability of iSPEX in the fast-moving smartphone market was limited by its need for device-specific calibrations.

Consumer unmanned aerial vehicles (UAVs) with RGB cameras have similarly become common scientific instruments. They provide a low-cost, high-resolution replacement for, or complement to, satellite and airplane imagery, especially for environmental monitoring \cite{Manfreda2018UAS, Ruwaimana2018droneadvantages, Rasmussen2016UAVvegetation, Tauro2016surfaceflow, Berra2015UAVs, Flynn2014UAVaquaticvegetation}. UAV data are increasingly being integrated with data from other platforms, such as satellites \cite{Gray2018UAVandsatellite}.

However, few scientific consumer camera projects progress past a proof-of-concept on a handful of camera models, which often become obsolete within two years, particularly in the constantly shifting smartphone market. This severely limits the sustainability of projects that require calibrations specific to each camera model. Difficulties in upscaling and future-proofing such calibrations are an oft cited constraint on the combination of multiple camera models \cite{Igoe2018ozone, Rateni2017smartphonefoodreview, Whiteson2016cosmicrays, Levin2016fluoride, Snik2014ispex, Nguyen2014raw2raw, Goddijn2009fundamentals}. Further complications are introduced by the lack of control over camera hardware and software parameters such as focus and white balance \cite{Ruwaimana2018droneadvantages, Haenel2018DSM, Kim2017smartphonebioluminescence, Zhang2016GFresnel, Snik2014ispex, Skandarajah2014microscope}. For example, the dominant smartphone operating systems, Android and iOS, only introduced support for unprocessed (RAW) imagery as recently as 2014 (Android 5.0 `Lollipop') and 2016 (iOS 10). Previously, third-party developers could only use JPEG data, which introduce a number of systematic errors due to their lossy compression and bit-rate reduction \cite{SanchezdeMiguel2019artificiallight, Leeuw2018hydrocolor, Igoe2018ozone, Rateni2017smartphonefoodreview, Zhang2016GFresnel, Snik2014ispex, Skandarajah2014microscope, Bongiorno2013spectralresponse, Kreuter2009allskypolarimetry}. Other common problems in consumer camera data include nonlinearity and the gamma correction \cite{SanchezdeMiguel2019artificiallight, Leeuw2018hydrocolor, Coburn2018remotesensingcameras, Ding2018mHealth, Xu2018communication, Wang2017herbicide, Manakov2016evaluation, Novoa2015WACODI, Sumriddetchkajorn2014chlorine, Snik2014ispex, Skandarajah2014microscope, Bongiorno2013spectralresponse, Charriere2013rgbmicroscope, Kreuter2010skypolarization, Kreuter2009allskypolarimetry, Lebourgeois2008commercialcameras, Cescatti2007canopycamera}, electronic and thermal noise \cite{Igoe2018ozone, Kim2017smartphonebioluminescence, Turner2017UVB, Pagnutti2017PiCamera, Whiteson2016cosmicrays, Berra2015UAVs, Igoe2014dark}, and highly variable (between camera models) spectral response functions which are not provided by manufacturers \cite{SanchezdeMiguel2019artificiallight, Leeuw2018hydrocolor, Coburn2018remotesensingcameras, Rateni2017smartphonefoodreview, McCracken2016resourcelimited, Berra2015UAVs, Javorsek2015nikonspectra, Nguyen2014raw2raw, Jiang2013spectralsensitivity, Bongiorno2013spectralresponse, Goddijn2009fundamentals}. These factors limit the accuracy of radiometric measurements done with consumer cameras by introducing systematic errors. Furthermore, the accuracy of color measurements and their conversion to standard measures, such as the CIE 1931 XYZ and CIELAB color spaces, is limited by distortions in the observed colors \cite{Kordecki2017vignettingSPIE} and differences in spectral response functions \cite{Nguyen2014raw2raw, Jiang2013spectralsensitivity, Bongiorno2013spectralresponse, Cheung2005multispectralcharacterization}.

Extensive (spectro-)radiometric calibrations of consumer cameras are laborious and require specialized equipment and are thus not commonly performed \cite{Manakov2016evaluation, Darrodi2015spectralsensitivity, Jiang2013spectralsensitivity}. A notable exception is the spectral and absolute radiometric calibration of a Raspberry Pi 3 V2 webcam by Pagnutti et al. \cite{Pagnutti2017PiCamera}, including calibrations of linearity, exposure stability, thermal and electronic noise, flat-field, and spectral response. Using this absolute radiometric calibration, digital values can be converted into SI units of radiance. However, the authors noted the need to characterize a large number of these cameras before the results could be applied in general. Moreover, certain calibrations are device-dependent and would need to be done separately on each device. Spectral and radiometric calibrations of seven cameras, including the Raspberry Pi, are given in \cite{Coburn2018remotesensingcameras}. These calibrations include dark current, flat-fielding, linearity, and spectral characterization. However, for the five digicams included in this work, JPEG data were used, severely limiting the quality and usefulness of these calibrations, as described above.

Spectral characterizations are more commonly published since these are vital for quantitative color analysis. Using various methods, the spectral responses of several Canon \cite{SanchezdeMiguel2019artificiallight, Coburn2018remotesensingcameras, Manakov2016evaluation, Berra2015UAVs, Jiang2013spectralsensitivity, Bongiorno2013spectralresponse, Zhao2009spectralsensitivity, Lebourgeois2008commercialcameras}, Nikon \cite{SanchezdeMiguel2019artificiallight, Manakov2016evaluation, Darrodi2015spectralsensitivity, Javorsek2015nikonspectra, Jiang2013spectralsensitivity, Bongiorno2013spectralresponse, Zhao2009spectralsensitivity, Sigernes2009absolutesensitivity, Goddijn2009fundamentals}, Olympus \cite{Coburn2018remotesensingcameras, Jiang2013spectralsensitivity, Bongiorno2013spectralresponse}, Panasonic \cite{Berra2015UAVs}, SeaLife \cite{Goddijn2009fundamentals}, Sigma \cite{Darrodi2015spectralsensitivity}, and Sony \cite{SanchezdeMiguel2019artificiallight, Coburn2018remotesensingcameras, Berra2015UAVs, Jiang2013spectralsensitivity, Zhao2009spectralsensitivity} digital cameras (digicams), as well as a number of smartphones \cite{Leeuw2018hydrocolor, Jiang2013spectralsensitivity}, have been measured. Direct comparisons between >2 different camera models are given in \cite{SanchezdeMiguel2019artificiallight, Leeuw2018hydrocolor, Coburn2018remotesensingcameras, Berra2015UAVs, Jiang2013spectralsensitivity}. Common features include the peak response wavelengths for the RGB color filters, approximately 600, 520, and 470 nm, respectively, as well as a roughly Gaussian profile around the peak. Differences are found especially in the wings, notably the locations of secondary peaks and near-infrared (NIR) and ultraviolet (UV) cut-off wavelengths. These may cause significant differences in observed colors between cameras \cite{Nguyen2014raw2raw, Jiang2013spectralsensitivity}, especially for narrow-band sources.

Camera calibrations in the literature are often limited to a small number of cameras or properties, either to narrow down the scope or because of limitations in time and equipment. Furthermore, calibration data are published in varying formats, locations, and quality, complicating their use by others. Standardized formats exist, such as those for vignetting, bias, and color corrections described in Adobe's digital negative (DNG) standard \cite{DNG1.4.0.0}, but have seen limited adoption. The European Machine Vision Association (EMVA) standard 1288 \cite{EMVA1288} for characterization of cameras is extremely thorough, but has also seen limited adoption due to the high-end equipment required \cite{Manakov2016evaluation} and its scope simply being too broad for many practical purposes. Similarly, standardized data sets or databases, for example containing spectral response curves \cite{Darrodi2015spectralsensitivity, Jiang2013spectralsensitivity}, have been created but these are limited in scope and, again, adoption. To our knowledge, no widely adopted standardized methodology or centralized database containing spectral and radiometric calibration data for consumer cameras has been created thus far.

In this work, we present a standardized methodology for the calibration of consumer cameras and a database, SPECTACLE (Standardised Photographic Equipment Calibration Technique And CataLoguE), containing calibration data for the most popular devices. The calibration methodology is focused on simplicity and facilitating measurements by non-experts and those lacking expensive equipment, similarly to \cite{Manakov2016evaluation} but with a broader scope including software, optics, and sensor characteristics. The database is designed with openness and sustainability in mind, focusing on community contributions. Furthermore, we strive to follow previously existing standards, such as DNG \cite{DNG1.4.0.0} and EMVA 1288 \cite{EMVA1288}, where practical. Our focus is on radiometric and photometric measurements but these calibration data can equally be used for color science purposes, in particular to convert between color spaces using the measured spectral response curves. We stress that we have no financial nor commercial interests in consumer cameras, and any comparison between devices is purely scientific. The aim of our standardized methodology and the SPECTACLE database is merely to simplify the use of data from consumer cameras, not to cast judgment on their quality.

Sect.~\ref{s:cameras} contains an overview of hardware and software trends in consumer cameras. We present the standardized calibration methodology in Sect.~\ref{s:methods}. Sect.~\ref{s:results} contains results from its application to several popular cameras and a description of the SPECTACLE database. Finally, in Sect.~\ref{s:discussion} we compare our findings with previous work and discuss future perspectives.

\section{Trends in consumer cameras} \label{s:cameras}

Consumer cameras can be divided into four categories, namely smartphones, UAVs, digicams (DSLR and mirrorless), and webcams. Despite serving very diverse purposes, these cameras share common characteristics and can be calibrated with the same methods.

CMOS-based sensors now dominate the consumer camera market \cite{Crocombe2018spectroscopy}. These are often not produced in-house by camera manufacturers, but acquired from external parties, such as Sony and Samsung. Different cameras often use the same sensor, such as the Sony IMX298 which is used in 12 smartphone models from 10 different manufacturers.

Most color cameras use Bayer filters, on-chip RGB filters arranged in a checkerboard pattern, with two green pixels (G and G$_2$) for every red or blue one \cite{Bayer1975}. The spectral responses of these filters differ strongly between cameras and are further modified by fore-optics \cite{Nguyen2014raw2raw}. Alternate pixelated filter arrangements exist, intended for example to reduce aliasing, but with little adoption so far \cite{Lin2016chroma}. Data from the separate RGBG$_2$ pixels can be recombined through a process known as demosaicing to retrieve an image with interpolated RGB values for each pixel. Many different schemes exist for this \cite{Lin2016chroma}, ranging from simple bilinear interpolation \cite{Pagnutti2017PiCamera, Kordecki2017vignettingSPIE, Berra2015UAVs} to complex computational methods \cite{Menon2007demosaicking}. Consumer camera software often includes proprietary demosaicing algorithms \cite{Skandarajah2014microscope, Lebourgeois2008commercialcameras} which may introduce complex, unpredictable effects. Depending on their implementation, demosaicing schemes typically mix data from different filters and remove their mutual independence, leading to undesirable cross-feed effects \cite{Leeuw2018hydrocolor, Pagnutti2017PiCamera}. In any case, the added data are fully synthetic and thus do not offer any new radiometric information. It is thus preferable for radiometric applications to treat the RGBG$_2$ images completely independently \cite{Lebourgeois2008commercialcameras} and demosaic data for visualization purposes \cite{Pagnutti2017PiCamera} only.

As discussed previously, the most commonly used digital file formats are JPEG (or JPG) and RAW. In both formats, data are saved on a pixel-by-pixel basis in analog-digital units (ADU). ADU are alternately referred to as digital numbers (DN) in the literature, but in this work we will use the ADU nomenclature. JPEG (ISO 10918) is based on lossy spatial compression and downsampling to 8-bit values, optimal for small file sizes while maintaining aesthetic qualities. Due to camera-specific processing and compression artefacts, JPEG images lose information and are not recommended for quantitative analysis \cite{SanchezdeMiguel2019artificiallight, Leeuw2018hydrocolor, Igoe2018ozone, Rateni2017smartphonefoodreview, Zhang2016GFresnel, Snik2014ispex, Skandarajah2014microscope, Bongiorno2013spectralresponse, Kreuter2009allskypolarimetry, Lebourgeois2008commercialcameras}. While standardizations exist, such as the standard Red Green Blue (sRGB) color space and gamma curve \cite{DNG1.4.0.0}, these are not strictly adhered to and cannot be assumed in data processing \cite{Finlayson2016spectralsensitivity}. Conversely, RAW files contain relatively unprocessed sensor output, intended for manual post-processing. One factor complicating the reduction of RAW data is their mosaiced nature, due to which they must be demosaiced or treated as multiple independent images, as discussed above. Despite these complications, their unprocessed nature makes RAW data highly preferable for scientific purposes \cite{Leeuw2018hydrocolor, Rateni2017smartphonefoodreview, Pagnutti2017PiCamera, Nguyen2014raw2raw, Bongiorno2013spectralresponse, Lebourgeois2008commercialcameras}.

Available camera controls generally include focus, exposure time, ISO speed (sensitivity), and aperture. Focus and aperture are changed by physical movement of camera optics, though most webcams and smartphones only allow a single, fixed aperture. ISO speed is set by changing the camera gain, through analog amplification or digital processing. Analog amplification involves varying the gain of the CMOS amplifiers, which can be done on the level of individual pixels. Conversely, digital gain is implemented in post-processing by simply re-scaling and interpolating measured digital values. Since ISO speed is a measure of the overall sensitivity of the camera, including fore-optics, each camera (and possibly each pixel) has a unique relation between ISO speed and gain. Finally, exposure time may be set by a physical shutter (common in digicams) or an electronic one (common in smartphones). Other parameters like white balance only affect processed imagery and are not relevant to RAW photography.

Many cameras include some built-in calibrations, most notably for nonlinearity, dark current, and flat-fielding effects. Nonlinearity corrections are typically based on previously measured correction curves \cite{Zhang2011linearCMOS}. Dark current corrections (autodarking) are commonly done using unilluminated exposures or permanently dark pixels around the sensor. Finally, flat-fielding (specifically vignetting) is typically corrected using a pre-made correction map. A variety of methods for generating such maps exists, based for example on computational methods using regular photographs \cite{Kordecki2017vignettingSPIE, Silva2016vignetting, Goldman2010vignetting, Zheng2009vignetting, Kim2008radiometryvignetting}, simply averaging many exposures \cite{Lebourgeois2008commercialcameras}, and simply imaging white paper \cite{Yu2004vignetting}. These maps are typically parametrized, for which various methods also exist \cite{Pagnutti2017PiCamera, Kordecki2017vignettingSPIE, DNG1.4.0.0, Goldman2010vignetting, Zheng2009vignetting, Kim2008radiometryvignetting, Lebourgeois2008commercialcameras, Yu2004vignetting}, the simplest being the $\cos^4$ model, a combination of inverse square falloff, Lambert's law, and foreshortening \cite{Goldman2010vignetting}. Alternately, a pixel-by-pixel map of vignetting correction coefficients may be used. Such maps may be device-specific or generalized for a camera model. Notably, iOS-based smartphones use the seven-parameter parametrization described in the DNG standard \cite{DNG1.4.0.0} (see Sect.~\ref{ss:methods:flat}) while Android-based smartphones use pixel-by-pixel maps.

\subsection{Smartphones} \label{ss:cameras:smartphones}

The smartphone market has become remarkably homogeneous in recent years, with virtually all models using the slate form factor, featuring a large touch screen, few buttons, and a camera on either side of the device. The most popular smartphones are all iOS- or Android-based. Both these operating systems now support RAW photography using Adobe's DNG standard \cite{DNG1.4.0.0}, though not on all devices. Hardware properties are rarely released by manufacturers, and are instead often provided by reviewers through disassembly of the smartphone.

Smartphone cameras aim to reproduce the human eye and thus have similar optical properties \cite{Leeuw2018hydrocolor}. Sensors, most commonly from the Sony Exmor series, are compact with 12--16 megapixels and a diagonal of 5--8 mm. Some devices focus on high-resolution imagery with many pixels, while others are optimized for low-light conditions, with fewer but larger pixels.

Smartphones now increasingly have multiple rear cameras. These secondary cameras offer features such as different fixed focal lengths and higher sensitivity, for example with a different lens or a monochromatic sensor. All rear cameras are typically placed in a cluster at the top right or top center of the smartphone.

\section{Methods} \label{s:methods}

In this section we describe the standardized methods for the calibration of consumer cameras. We developed a custom data processing pipeline, implemented in Python scripts available on GitHub (\url{https://github.com/monocle-h2020/camera_calibration}) and iOS and Android apps (\url{https://github.com/monocle-h2020/spectacle_android} and \url{https://github.com/monocle-h2020/spectacle_ios}). 

Sect.~\ref{ss:methods:setup} describes the experimental setups and data processing used in calibration measurements. The methods used to characterize and calibrate the camera responses are given in Sects.~\ref{ss:methods:survey}--\ref{ss:methods:spectral}. Finally Sect.~\ref{ss:methods:calibration} describes how consumer camera data are converted into relative radiometric units using the previously described calibration measurements. These units provide a constant scale, independent of exposure parameters and individual device characteristics, for each camera model, a constant factor $K$ per model away from absolute radiometric units (W m$^{-2}$ sr$^{-1}$). Absolute radiometric calibration is outside the scope of this work.

\subsection{Experimental setup} \label{ss:methods:setup}

This section describes the setups used in our golden standard ground-truth experiments. Descriptions of do-it-yourself (DIY) calibration methods are given in the relevant sections. All images from all cameras were taken in RAW format; for the linearity measurements, simultaneous RAW and JPEG images were taken for comparison. As discussed in Sect.~\ref{s:cameras}, demosaicing schemes introduce synthetic data and undesirable cross-feed effects. For this reason, in our data reduction the RAW images were split into separate RGBG$_2$ images which were analyzed individually \cite{Lebourgeois2008commercialcameras}. Multiple images were taken and stacked for each measurement to improve the signal-to-noise ratio (SNR). On smartphones, the aforementioned iOS and Android apps were used to control the camera and automatically take multiple exposures. Exposure settings, including ISO speeds and exposure times were obtained from camera controls where possible, since EXIF metadata values for these were found (Sect.~\ref{ss:results:survey}) to be unreliable.

\begin{figure}[ht]
	\centering
    \includegraphics[width=\textwidth]{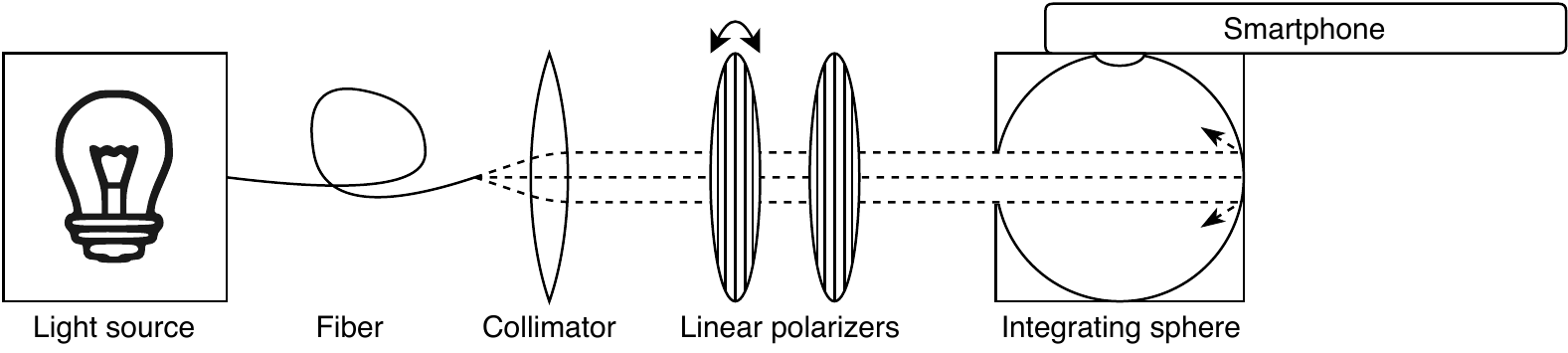}
    \caption{Setup used to measure linearity, ISO-gain relations, and inter-pixel gain variations on smartphones. The first linear polarizer was rotatable, the second fixed. Smartphones were placed with their camera flush against the view-port at the top of the integrating sphere.}
    \label{f:setup_linearity_polarizers}
\end{figure}

The setup for measuring linearity, ISO-gain relations, and inter-pixel gain variations on smartphones is shown in Fig.~\ref{f:setup_linearity_polarizers}. A halogen light source (OceanOptics HL-2000-LL) was used, specified by the manufacturer to be stable to 0.15\% peak-to-peak and drift <0.3\% per hour after a warm-up of 10 minutes. Its light was fed into an optical fiber (Thorlabs M25L02) and collimated using two lenses (Thorlabs AC254-030-A with $f = 30$ mm and AC508-200-A with $f = 200$ mm). Two linear polarizers (both Thorlabs LPVISE100-A, with an extinction ratio $\geq$495 from 400-700 nm), the first rotatable and the second fixed, were used to attenuate the light beam entering an integrating sphere (Thorlabs IS200). Using Malus's law ($I = I_0 \cos^2 \theta$), the rotation angle between the polarizers could be used to calculate the attenuation. A calibration detector was not necessary since all experiments done with this setup involve relative measurements only. Malus's law was first fitted to a series of exposures over the entire rotation range to determine the reference angle. The rotation angle of the first polarizer could be determined visually up to 2\textdegree{} precision, giving a typical uncertainty on the attenuated intensity of 2.5\%. Finally, smartphones were placed on top of the integrating sphere, flush against the view-port. The farthest possible focus was used (infinity on Android devices, an arbitrary number on iOS). All experiments done with this setup involved analysis on the individual pixel and (broad-band) filter level, without any spatial averaging. Because of this, differences in illumination due to spectral dependencies in the polarizer throughput or the integrating sphere output did not affect any of the experiments.

The linear polarizers can be replaced with alternate methods for attenuation, such as neutral density filters. Attenuation can also be replaced completely by varying exposure times instead, though physical attenuation may be more precise \cite{Pagnutti2017PiCamera}. The integrating sphere may be replaced by another diffuse surface, such as a Spectralon target. If sufficiently wide, the light beam may also be shone directly onto the sensor; such a setup was used for digicams, with the digicam in place of the collimator in Fig.~\ref{f:setup_linearity_polarizers} at a sufficient distance to completely illuminate the sensor. This was done to simplify the alignment process since our digicams had large physical CMOS sizes. Since all measurements were done on the individual pixel level, they were not affected by the added differences in illumination.

Bias, read-out noise, and dark current were measured on all devices by placing the camera flush against a flat surface (such as a table), pointing down, in a dark room. The setups for flat-fielding and spectral characterization are described in Sects.~\ref{ss:methods:flat} and \ref{ss:methods:spectral}, respectively.

\subsection{General properties} \label{ss:methods:survey}

General hardware and software properties were retrieved from official specifications and community reviews. A survey across these provided an overview of basic physical and optical parameters of cameras. On Android smartphones, the Camera2 API provides ample information on such parameters, facilitating automatic data collection using an app. 

The retrieved device properties included the camera type, manufacturer, product code and internal identifiers, release year, the number of cameras (for smartphones), camera module identifiers, and CMOS sensor models. Sensor properties included physical size, pixel pitch, resolution, orientation with respect to the device, color filter pattern, and bit depth. Camera optic properties included focal length, $f$-number, neutral density filters (for high-end smartphones), and a vignetting model if available. Finally, software and firmware properties included supported software versions, RAW and JPEG support, estimated bias value (see Sect.~\ref{ss:methods:bias_ron}), ISO speed range, exposure time range, and the active part of the sensor (accounting for dark pixels, see Sect.~\ref{ss:methods:dark}).

\subsection{Linearity} \label{ss:methods:linearity}
Sensor linearity was quantified by measuring the camera response to varying exposures, either by attenuating a light source or by varying the exposure time, as discussed in Sect.~\ref{ss:methods:setup}. We used the setup shown in Fig.~\ref{f:setup_linearity_polarizers} with two linear polarizers to attenuate the light for smartphones, since exposure times on those are not completely trustworthy (Sect.~\ref{ss:results:survey}). Conversely, for digicams, exposure times are reliable \cite{Manakov2016evaluation, Darrodi2015spectralsensitivity} and thus were used instead of physical attenuation to simplify the setup. A third method, varying the physical aperture, changes the distribution of light on the sensor \cite{Goldman2010vignetting} and thus cannot be used to measure linearity.

Two common types of nonlinearity exist, either across the entire intensity range or only at high intensities. The former is common in JPEG imagery due to the gamma correction \cite{Skandarajah2014microscope, Lebourgeois2008commercialcameras} while the latter is expected in both JPEG and RAW data. We only investigated the former since it has the largest impact on data quality, as described in Sect.~\ref{s:intro}. Nonlinearity at high intensities is easily negated by discarding data above a threshold value; we use a threshold of $\geq$95\% of the maximum digital value.

The linearity of each pixel was expressed through the Pearson correlation coefficient $r$, a measure of the linear correlation between intensity and camera response. Pixels were analyzed individually to negate differences in illumination and vignetting effects (Sect.~\ref{ss:methods:flat}). Simulated responses of a perfectly linear camera with a mean error of 5\% in the incoming intensity simulating, for example, errors in exposure parameters or polarizer alignment in the setup described in Sect.~\ref{ss:methods:setup}, as well as Poisson noise ($\sigma_N = \sqrt{N}$) and 10 ADU read noise in the response were analyzed. This included simulated measurements at 15 different exposures, averaged over 10 images per exposure. These simulated data resulted in a mean value of $r = 0.996 \pm 0.002$ and a lower 0.1 percentile $P_{0.1}(r) = 0.985$. To account for unforeseen measurement errors, we set the cut-off for linearity at $r \geq 0.980$.

Additionally, the JPEG data were compared to sRGB-like profiles to determine whether gamma inversion \cite{Novoa2015WACODI} is possible. The sRGB-like profiles are described by Eq.~\eqref{e:sRGB}, with $J_\mathcal{C}$ the JPEG response (0--255) in band $\mathcal{C}$, $n$ a normalization factor, $\gamma$ the gamma correction factor and $I$ the incoming intensity in arbitrary units. The JPEG response of each pixel was individually fit to Eq.~\eqref{e:sRGB} with $n$ and $\gamma$ as free parameters. Additionally, profiles with standard $\gamma$ values (2.2 and 2.4 \cite{Novoa2015WACODI}) were fit to the JPEG data (with $n$ free) to determine the accuracy of these standards.

\begin{equation} \label{e:sRGB}
    J_\mathcal{C} = 255 \times \begin{cases}
    12.92 n I                      & \text{if } n I < 0.0031308 \\
    1.055 (n I)^{1/\gamma} - 0.055 & \text{otherwise}
    \end{cases}
\end{equation}

\subsection{Bias \& read-out noise} \label{ss:methods:bias_ron}

Bias (or `black level') and read-out noise (RON) were measured by stacking short dark exposures. The bias and RON in individual pixels are given by the mean and variance, respectively, of their values in each stack. Many (>50) images per stack are required to distinguish bias variations from RON. Temporal variations were probed by repeating this process several times. While EXIF metadata often contain a bias value, this is only an estimate and should be validated by measurement.

\subsection{Dark current} \label{ss:methods:dark}

Dark current (thermal noise) was measured by taking dark exposures with different lengths and fitting a linear relation between exposure time and camera response to determine the dark current in ADU s$^{-1}$. For cameras that have autodarking (see Sect.~\ref{s:cameras}), the residual dark current was characterized instead. Depending on the autodark precision, the exposure-response relation may be non-linear in this case.

\subsection{ISO speed} \label{ss:methods:iso}

The relation between camera sensitivity and ISO speed was measured by taking identically exposed images at different ISO speeds. These were bias-corrected and pixel values were divided by those at the lowest ISO speed. A relation between ISO speed and normalization factor was then fitted. Like the linearity measurements (Sect.~\ref{ss:methods:linearity}), this was done individually per pixel to negate illumination differences and vignetting effects.

This relation may be any combination of linear and constant functions, depending on the implementation of ISO speed ratings. Linear relations correspond to analog gain, while digital gain may result in linear or constant relations, as described in Sect.~\ref{s:cameras}.

\subsection{Gain variations} \label{ss:methods:gain}

Inter-pixel and inter-filter gain variations were characterized using the mean-variance method \cite{McLean2008electronicimaging}, which exploits the Poissonian nature of photo-electrons in a sensor. We applied this method to individual pixels rather than averaging over the sensor, to measure inter-pixel variations and remove the need for flat-fielding prior to this calibration. The response of a digital camera to incoming light is given by Eq.~\eqref{e:mean}, with $M$ the mean response in ADU, $I$ the exposure in photo-electrons, $D$ the dark current in e$^-$, $B$ the bias in ADU, and $G$ the gain in ADU/e$^-$. Both $I$ and $D$ are integrated over the exposure time.

\begin{equation} \label{e:mean}
M = I G + D G + B
\end{equation}

The variance in the response of a pixel is a combination of shot noise on the photo-electrons and dark current, and read noise. The shot noise follows a Poissonian distribution with a standard deviation $\sigma_I = \sqrt{I}$ and thus a variance $V_I = I$. The total variance in the response is expressed in Eq.~\eqref{e:var}, with $V$ the variance in ADU$^2$ and $RON$ the read noise in ADU.

\begin{equation} \label{e:var}
V = I G^2 + D G^2 + RON^2
\end{equation}

After correcting for bias and dark current, and assuming $D G^2$ is negligible, a linear relation between mean and variance is found, shown in Eq.~\eqref{e:gain}.

\begin{equation} \label{e:gain}
V = G M_\text{cor} + RON^2
\end{equation}

Equation~\eqref{e:gain} was fitted to mean and variance values from several image stacks taken under different illumination conditions. Within each stack, all images were exposed identically, while the illumination varied between stacks. A large amount of data (>10 stacks of >50 images each) was necessary to constrain the fitted gain values sufficiently (typical relative errors in individual pixels <15\%). ISO normalization functions derived in Sect.~\ref{ss:methods:iso} may be used to extrapolate measured values to different ISO speeds.

\subsection{Flat-field correction} \label{ss:methods:flat}

Flat-fielding was performed by imaging a uniform light source. Unlike telescopes, most consumer cameras have fields-of-view (FoV) too large to use the twilight sky for this. Instead, a large integrating sphere was used to create an isotropic light field, as described in \cite{Pagnutti2017PiCamera}. We used a LabSphere HELIOS USLR-D12L-NMNN lit by three halogen lamps with a specified luminance uniformity of $\pm$1.0\%, sequentially placing each camera before its aperture.

Any significant chromatic differences in the flat-field response were measured automatically, since all filters were exposed simultaneously. The RGBG$_2$ images were split out and each normalized to their maximum value, then recombined and smoothed with a Gaussian filter ($\sigma = 10$ pixels); both the individual RGBG$_2$ images and the recombined image were analyzed. Since vignetting, often the dominant flat-field component, is caused by the camera aperture, the flat-field response changes and must be measured again when varying the aperture \cite{Goldman2010vignetting}.

Vignetting can be parametrized in a number of different ways, as discussed in Sect.~\ref{s:cameras}. For consistency, we used the DNG seven-parameter ($k_0 \ldots k_4, \hat{c}_x, \hat{c}_y$) model, also used internally in iOS smartphones, for the flat-field correction factor $g(x,y)$, expressed in Eq.~\eqref{e:vignetting}, with $r$ the normalized Euclidean distance from pixel $(x,y)$ to the optical center $(\hat{c}_x, \hat{c}_y)$.

\begin{equation} \label{e:vignetting}
g(x,y) = 1 + k_0 r^2 + k_1 r^4 + k_2 r^6 + k_3 r^8 + k_4 r^{10}
\end{equation}

Three simpler, alternate methods were also tested. The first involved imaging an overcast sky, the second imaging the sun with a piece of paper taped onto the camera as a diffuser similarly to the Hukseflux Pyranometer app (\url{http://www.hukseflux.com/product/pyranometer-app}). For the final method, the camera, again with a paper diffuser, was held flush against a computer monitor displaying a white screen, somewhat similarly to \cite{Manakov2016evaluation}. In all three cases, the camera was dithered and rotated 360$^\circ$ during measurements to average out anisotropies. Data from all three methods were processed in the same way as the integrating sphere data, to compare their efficacy.

\subsection{Spectral response} \label{ss:methods:spectral}

The spectral response of a camera, which is a product of the individual spectral responses of its fore-optics, filters, and sensor, was measured in two ways. The first method, using a monochromator, is simple processing-wise as the data are simply a series of images at different wavelengths with known intensities \cite{Pagnutti2017PiCamera, Berra2015UAVs, Sigernes2009absolutesensitivity}. It also allows for the measurement of inter-pixel variations in spectral response. The second, a spectrometer add-on such as iSPEX \cite{Snik2014ispex}, is more accessible than monochromators but its spectral data are more difficult to calibrate and process.

We used a double monochromator (OL 750-M-D) at the NERC Field Spectroscopy Facility to scan a wavelength range of 390-700 nm. This wavelength range was chosen because no significant response was found outside it on any of the test cameras. The effective spectral resolution (half bandwidth) of the monochromator was 4 nm, calculated from the grating (1200 grooves/mm) and slits (2.5 mm entrance/exit and 5.0 mm central slit) used. The wavelength range was critically sampled at 2 nm intervals. A laser-driven light source (Energetiq EQ-99X) was used, and its spectral output calibrated using a silicon photodiode (Gooch \& Housego OL DH-300C with a Hamamatsu S1337-1010BQ sensor). The system was NIST-traceably calibrated in 2012 and is described in more detail in \cite{Berra2015UAVs}.

Spectral characterization was also done using a modified (removing polarizers and retarders) iSPEX add-on \cite{Snik2014ispex}. iSPEX has a slit consisting of two parts, one 0.4 mm (`broad') and the other 0.17 mm (`narrow') wide and a 1000 grooves/mm transmission grating foil (Edmund Optics \#52-116). Using this foil, a similar spectrometer can be built for any other camera.

The reflection of sunlight on a piece of white paper was measured using the iSPEX on an iPhone SE. iSPEX projects a spectrum onto the sensor, so the pixel responses must be corrected for bias, dark current, and flat-field to obtain a quantitative spectrum. The 436.6, 544.5, and 611.6 nm spectral lines of a commercial fluorescent lamp were used for the wavelength calibration, fitting a quadratic relation between pixel position and wavelength. A stray light correction was done by subtracting the mean pixel value per column above and below the spectrum from the narrow and broad slit spectra, respectively. Two theoretical reference spectra were used to normalize the observed spectra, namely a 5777 K black body (approximating the Sun) and a diffuse solar irradiance spectrum generated using the Simple Model for the Atmospheric Radiative Transfer of Sunshine (SMARTS2) \cite{Gueymard2001SMARTS2, Gueymard1995SMARTS2} and smoothed to the 5 nm resolution of narrow-slit iSPEX spectra. For the latter, the location and time of the iSPEX measurements as well as the built-in urban aerosol and ground albedo models were used instead of default parameters. The models differed significantly (RMS 34\%) due to the diffuse sky irradiance factored into the SMARTS2 model. Finally, the observed spectra were corrected for the transmission of the iSPEX optics, determined by measuring the zero-order transmission using a halogen lamp and spectrometer (OceanOptics HL-2000-LL and USB2000+, respectively).

Instead of the sun, a previously calibrated commercial lamp may be used. For example, the LICA-UCM database (\url{https://guaix.fis.ucm.es/lamps_spectra}) contains spectra of common commercial lamps which can be used as standard light sources for spectral response measurements \cite{Tapia2018LICALamps}. This method has the advantage of independence from weather conditions and higher reproducibility compared to solar measurements. Combined with the new version of iSPEX we are currently developing, featuring a universal smartphone hardware interface, this enables volunteer measurements of smartphone camera spectral responses.

The spectral curves $R_\mathcal{C} (\lambda)$ thus derived were normalized to the global maximum transmission in all bands and used for calibration of spectral measurements and in the radiometric correction of imaging data (Sect.~\ref{ss:methods:calibration}) to calculate effective spectral bandwidths $\Lambda_\mathcal{C}$. These are defined as $\Lambda_\mathcal{C} = \int_\mathcal{C} R^\prime_\mathcal{C}(\lambda) \text{d}\lambda$, with $R^\prime_\mathcal{C}(\lambda)$ the spectral response $R_\mathcal{C} (\lambda)$ normalized to the maximum in band $\mathcal{C}$ \cite{Pagnutti2017PiCamera, Schott2007remotesensingimagechain}. This integral was calculated using the composite trapezoid method, implemented in the NumPy function \verb|numpy.trapz|.

\subsection{Relative radiometric calibration} \label{ss:methods:calibration}

The calibrations described in the previous section are used to convert digital values to radiance. Following the methods described in \cite{Pagnutti2017PiCamera, Schott2007remotesensingimagechain, Beisl2006absolutespectroradiometric, Brown2001absoluteradiometric}, a digital value $M$ (in ADU) in band $\mathcal{C}$ (RGBG$_2$ for Bayer filters) can be converted to effective radiance $L_\mathcal{C}$, in units of W m$^{-2}$ sr$^{-1}$. 

Since absolute radiometric calibration is outside the scope of this work, we instead determined the relative effective radiance $L^\prime_\mathcal{C} = L_\mathcal{C} / K$, in relative radiometric units (RRU) m$^{-2}$ sr$^{-1}$, with $K$ an extra factor accounting for the absolute quantum efficiency and transmission of the lens. Measuring these requires a previously calibrated light source with a known radiance. 

The expression for converting $M$ to $L^\prime_\mathcal{C}$ is given in Eq.~\eqref{e:radiometry}. The advantage of the piece-wise calibration given in Eq.~\eqref{e:radiometry} over a black-box approach containing all calibration components is its adaptability when a small subset of parameters are changed, such as due to firmware updates or manufacturing changes. This way, calibration data can be re-used rather than requiring a full re-calibration with every change.

\begin{equation} \label{e:radiometry}
    L^\prime_\mathcal{C} = h c \frac{1}{A_d \Lambda_\mathcal{C}} g \left[\frac{4 (f\#)^2}{\pi \tau N}\right] (M - B - D \tau)
\end{equation}

First, the bias $B$ (ADU; Sect.~\ref{ss:methods:bias_ron}) and dark current $D \tau$ (with $D$ in ADU s$^{-1}$ and $\tau$ the exposure time in seconds; Sect.~\ref{ss:methods:dark}) are subtracted. Linearization of digital values \cite{Brown2001absoluteradiometric} is not necessary since we only used sufficiently linear pixels ($r \geq 0.980$; Sect.~\ref{ss:methods:linearity}).

Next, the image is corrected for the exposure parameters, dividing by the exposure time $\tau$, ISO speed normalization factor $N$ (Sect.~\ref{ss:methods:iso}), and aperture, approximated as $\pi/4 (f\#)^2$, with $f\#$ the $f$-number of the camera \cite{Pagnutti2017PiCamera}. This approximation causes a systematic error of 4\% at $f/2.0$ \cite{Pagnutti2017PiCamera}; for fixed-aperture systems like smartphones, this error is not relevant. For systems with adjustable apertures, an exact solution may be preferable if operating at very low $f$-numbers. These corrections yield a response in normalized ADU s$^{-1}$ sr$^{-1}$.

The third step is the flat-field correction. The response is multiplied by the flat-field correction $g$ (unitless; Sect.~\ref{ss:methods:flat}). The flat-fielding methods used here account for both optical and electronic variations in sensitivity, so a separate correction for inter-pixel gain variations (Sect.~\ref{ss:methods:gain}) is not necessary. Since absolute transmission and quantum efficiency were not measured, this step yields a response in relative counts s$^{-1}$ sr$^{-1}$, proportional to the number of photo-electrons s$^{-1}$ sr$^{-1}$.

Next, sensor properties are corrected for. The response is divided by the pixel size $A_d$ (m$^2$; Sect.~\ref{ss:methods:survey}) to give a response in relative counts s$^{-1}$ m$^{-2}$ sr$^{-1}$. It is then divided by the effective spectral bandwidth of band $\mathcal{C}$, $\Lambda_\mathcal{C} = \int_\mathcal{C} R^\prime_\mathcal{C}(\lambda) \text{d}\lambda$ (Sect.~\ref{ss:methods:spectral}).

Finally, the result is converted to a relative radiance by multiplying by a factor $h c$, with $h$ Planck's constant and $c$ the speed of light. This yields $L^\prime_\mathcal{C}$ in RRU m$^{-2}$ sr$^{-1}$.

For specific applications, Eq.~\eqref{e:radiometry} may be simplified or adjusted. For example, inter-pixel bias and dark current variations are typically negligible in bright conditions. In those cases, $B$ and $D$ may be approximated by constants, and inter-pixel variations incorporated in the error budget. For spectroscopic applications, a relative spectral radiance $L^\prime_{\mathcal{C}, \lambda}$ in RRU m$^{-2}$ sr$^{-1}$ nm$^{-1}$ is measured, which is not averaged over band $\mathcal{C}$. In this case, the energy per photon is simply $h c / \lambda$ and only the transmission at wavelength $\lambda$, $R_\mathcal{C}(\lambda)$ is relevant; furthermore, the result must be divided by the wavelength coverage of each pixel $\Delta \lambda$. This is expressed in Eq.~\eqref{e:radiometry_spectral}.

\begin{equation} \label{e:radiometry_spectral}
    L^\prime_{\mathcal{C}, \lambda} = \frac{h c}{\lambda} \frac{1}{A_d R_\mathcal{C}(\lambda) \Delta \lambda} g \left[\frac{4 (f\#)^2}{\pi \tau N}\right] (M_\lambda - B - D \tau)
\end{equation}

\section{Results} \label{s:results}
The methodology described in Sect.~\ref{s:methods} was applied to three iOS smartphones (Apple iPhone SE, 6S, and 7 Plus), two Android devices (Samsung Galaxy S6 and S8), one digicam (Nikon D5300), and one UAV camera (DJI Phantom Pro 4). This section contains an overview of results from these various calibration steps. Results for all devices are included in the SPECTACLE database further described in Sect.~\ref{ss:results:database}.

\subsection{General properties} \label{ss:results:survey}

General hardware and software properties were retrieved from the survey described in Sect.~\ref{ss:methods:survey}, with a specific focus on smartphones using the previously described Android app. Little variation was found in these general properties, especially for smartphones. For example, virtually all main cameras on smartphones have apertures of $f/2.4$--$f/1.5$, focal lengths of 3.8--4.5 mm, and sensors of 3.4--6.7 $\times$ 2.7--4.7 mm, giving fields-of-view (FoVs) of 60--75$^\circ$ $\times$ 45--55$^\circ$.

It was found from test images that EXIF metadata from some cameras are inaccurate. For example, the iPhone SE can use unrounded exposure times of $1/3.0$ s and $1/3.9$ s but records both as simply $1/3$ s in metadata. Assuming the recorded exposure time of $1/3$ s for a real exposure of $1/3.9$ s would lead to photometric errors up to 30\%. To counteract this, exposure parameters such as ISO speed and exposure time should be recorded separately from default EXIF metadata, for example with custom EXIF tags or extra files.

\subsection{Linearity} \label{ss:results:linearity}

The linearity of two smartphones (iPhone SE and Galaxy S8) and one digicam (Nikon D5300) was measured using the methods described in Sect.~\ref{ss:methods:linearity} and the setup described in Sect.~\ref{ss:methods:setup} and shown in Fig.~\ref{f:setup_linearity_polarizers}. The smartphones were analyzed using rotating linear polarizers while the D5300 was analyzed by varying exposure times. Simultaneous RAW and JPEG images were taken on each device (using the Fine JPEG setting on the D5300) to compare their responses. JPEG images were taken with a fixed white balance.

\begin{figure}[ht]
	\centering
    \includegraphics[width=\textwidth]{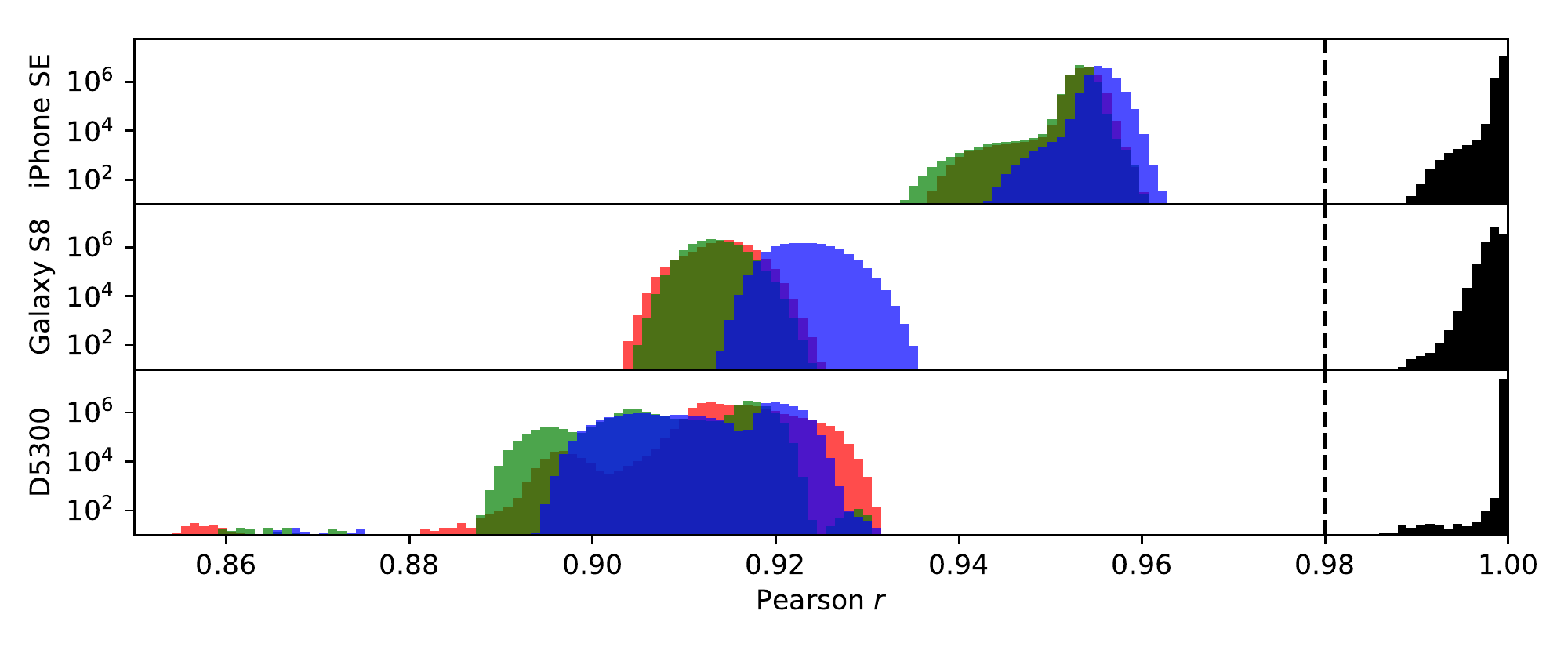}
    \caption{Histogram of Pearson $r$ coefficients for RAW (black, all filters combined) and JPEG (red/green/blue) responses. The $r \geq 0.980$ cut-off is shown with a dashed black line. The respective cameras are listed next to the vertical axis. Note the logarithmic vertical scale.}
    \label{f:linearity_hist}
\end{figure}

The Pearson $r$ coefficients of the RAW and JPEG responses of all pixels were calculated and their histograms are shown in Fig.~\ref{f:linearity_hist}. The JPEG responses of all pixels in all cameras were well below the linearity threshold ($r \geq 0.980$), showing again that JPEG data are highly nonlinear. Conversely, nearly all RAW responses were well within the bounds for linearity, with 99.9\% of $r$ values $\geq$0.997 (iPhone SE), $\geq$0.996 (Galaxy S8), and $\geq$0.999 (D5300). The Galaxy S8 was the only camera with RAW responses having $r < 0.980$, though only in 56 pixels.

\begin{figure}[ht]
	\centering
    \includegraphics[width=\textwidth]{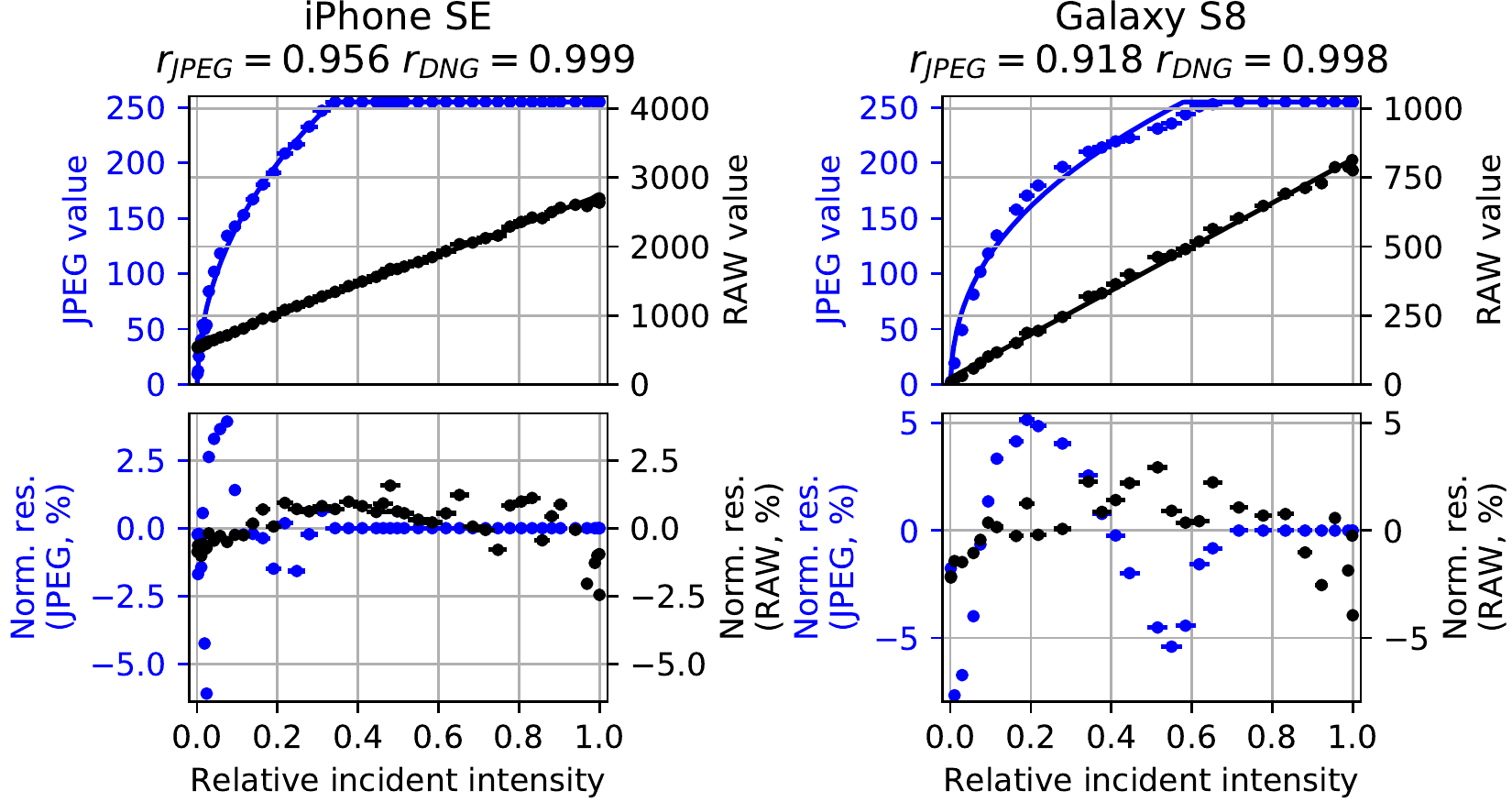}
    \caption{JPEG (blue, left vertical axis) and RAW (black, right axis) response of a single B pixel in the iPhone SE (left) and Galaxy S8 (right) rear cameras, under varying incident intensities. Each point represents the mean of a stack of 10 images at the same exposure. Vertical error bars are smaller than the dot size. The black and blue lines represent the best-fitting linear (RAW) and sRGB-like (JPEG) profiles, respectively. The lower row shows the residuals, normalized to the dynamic range.}
    \label{f:linearity_single_pixel}
\end{figure}

The JPEG and RAW responses of individual pixels in the iPhone SE and Galaxy S8 cameras are shown in Fig.~\ref{f:linearity_single_pixel}. The JPEG responses are visibly nonlinear ($r = 0.956, 0.918$) while the RAW responses are linear within measurement errors ($r = 0.999, 0.998$). Furthermore, the dynamic range of the JPEG data is much smaller than that of the RAW data. These differences highlight the advantages of RAW data.

\begin{figure}[ht]
	\centering
    \includegraphics[width=\textwidth]{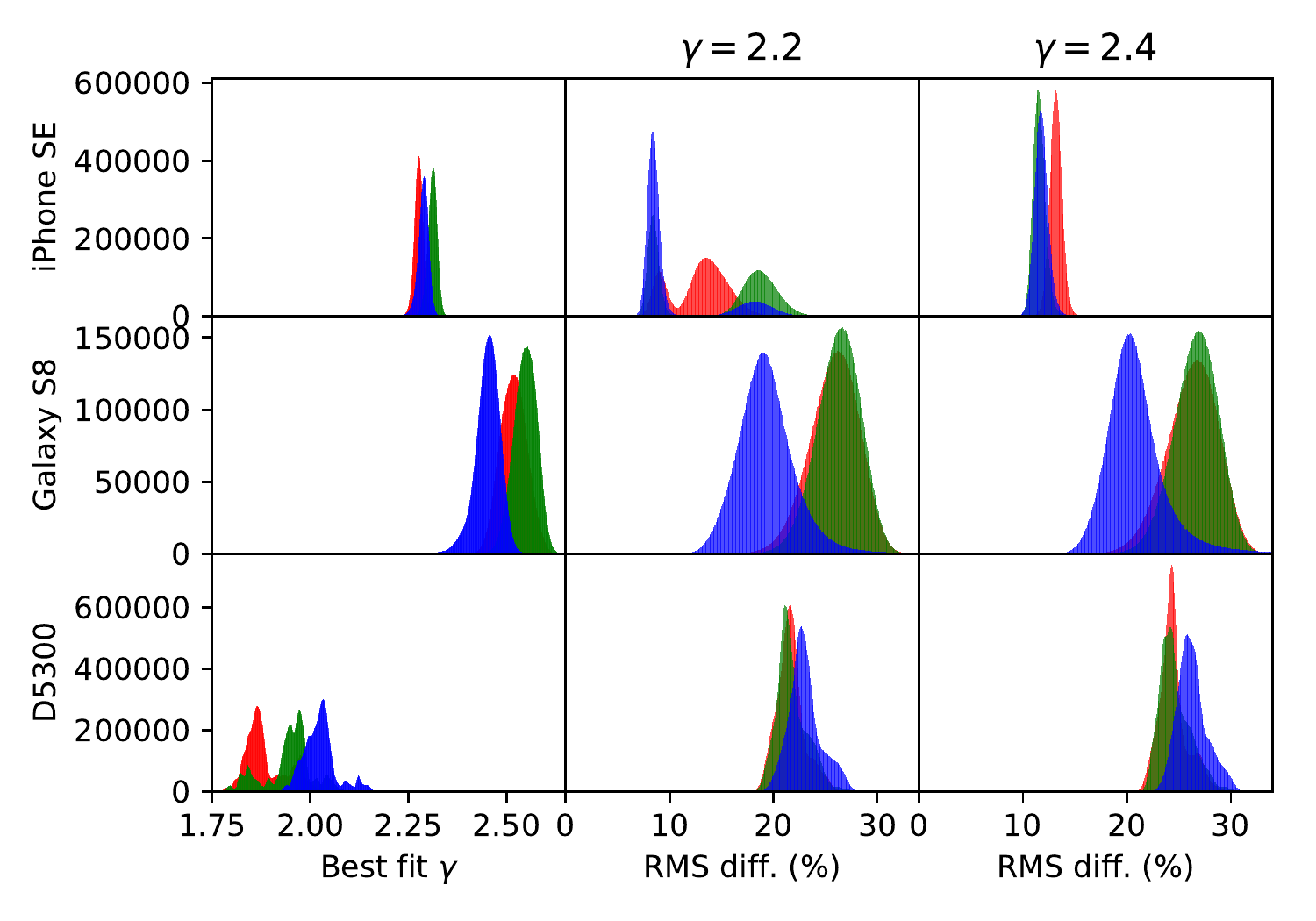}
    \caption{Histogram of best-fitting $\gamma$ and RMS relative difference between JPEG data and fit (for models with $\gamma = 2.2$ and $2.4$) in the RGB bands.}
    \label{f:jpeg}
\end{figure}

Finally, Fig.~\ref{f:jpeg} shows the best-fitting $\gamma$ for the JPEG response per pixel as well as the accuracy of two standard values ($\gamma = 2.2$ and $2.4$, expressed in RMS relative difference ($1 - \text{data}/\text{fit}$). Large inter-pixel, inter-filter, and inter-device differences in best-fitting $\gamma$ exist, indicating an sRGB gamma inversion with a single $\gamma$ value is not possible. Furthermore, the $\gamma = 2.2$ and $2.4$ models are both clearly very inaccurate for all cameras. For the $\gamma = 2.2$ and $2.4$ cases respectively, 99.9\% of pixels had RMS relative differences between observations and the sRGB model of >7\% and >10\% (iPhone SE), >13\% and >15\% (Galaxy S8), and >19\% and >21\% (Nikon D5300).

\subsection{Bias \& read noise} \label{ss:results:bias_ron}

Bias and read noise variations in four smartphone cameras (iPhone SE and 7 Plus, Galaxy S6 and S8), one digicam (Nikon D5300), and one UAV camera (Phantom Pro 4) were analyzed using the methods from Sect.~\ref{ss:methods:bias_ron}.

Bias values in all cameras deviated systematically from the EXIF values by <1 ADU on average, with standard deviations also <1 ADU. However, large outliers were found, such as some pixels in our Galaxy S6 which even saturated in bias frames. Phantom Pro 4 data are scaled up from 12-bit (its native bit depth) to 16-bit, increasing the observed bias variations. Scaled down to 12 bits, its bias variations are similar to those in the other cameras.

Typical observed RON values were distributed similarly to inter-pixel bias variations. The smartphones and D5300 show RON distributions consisting of one or two main components <3 ADU, which correlate with inter-pixel gain variations (Sect.~\ref{ss:results:gain}), and a long but shallow tail towards RON values >20 ADU. As with the bias variations above, the Phantom Pro 4 showed a comparatively high mean RON (14 ADU at ISO speed 100) in 16-bit (scaled-up) data but a comparable value (1.8 ADU) when scaled down to its native bit depth of 12 bits.

\begin{figure}[ht]
	\centering
    \includegraphics[width=\textwidth]{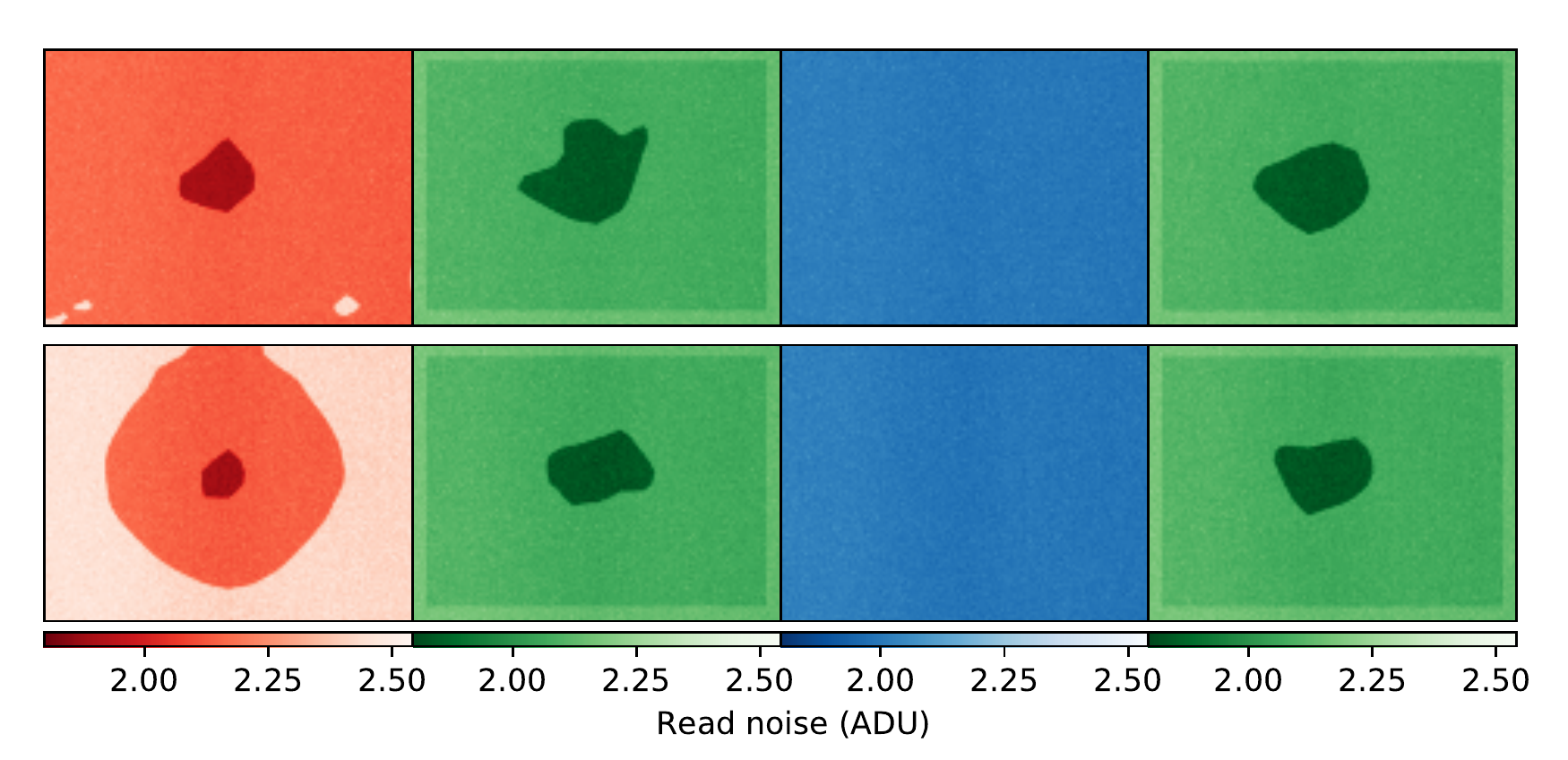}
    \caption{Read-out noise per pixel of two iPhone SE devices (top and bottom) at ISO speed 23, in the RGBG$_2$ filters from left to right. Darker colors correspond to lower read-out noise. A two-dimensional Gaussian filter ($\sigma = 5$ pixels) has been applied to better visualize large-scale variations. The G image shows similar patterns to Fig.~\ref{f:gain_map}.}
    \label{f:ron}
\end{figure}

Large-scale patterns in inter-pixel and inter-filter bias and RON variations were observed in several cameras, most prominently in the smartphones. Figure~\ref{f:ron} shows the RON per pixel in the sensors of two iPhone SE devices. The RON and bias patterns on each device are strongly correlated, suggesting a common origin. The RMS difference in bias between these two devices was 0.31 ADU, larger than the standard deviation on either device (0.24 and 0.21 ADU). The large-scale patterns persisted over time scales of months, indicating that they are systematic.

Both bias variations and RON decreased with ISO speed when normalized (Sect.~\ref{ss:methods:iso}). This may be a result of better amplifier or ADC performance at a higher gain. Similarly, large-scale patterns such as those in Fig.~\ref{f:ron} become less distinct at high ISO speeds.

Either a map of mean bias per pixel at a given ISO speed $B(x,y,ISO)$ or a mean value $B$ is used in Eq.~\eqref{e:radiometry}. For low-light applications such as dark-sky measurements \cite{Haenel2018DSM} or spectroscopy, a detailed map is necessary since a single `bad' pixel with an abnormally high output may cause a significant systematic error. Being manufacturing defects, bad pixels are in different locations even on two cameras of the same model, and thus a map is required for each device. Conversely, for bright conditions, the bias variations are not significant and thus a mean value can be used. Similarly, RON values can be incorporated in the error budget separately for individual pixels or using the RMS value as an ensemble estimate.

\subsection{Dark current} \label{ss:results:dark}

The methods described in Sect.~\ref{ss:methods:dark} were applied to two smartphones (iPhone SE and Galaxy S8) to measure their dark current properties. Both cameras have built-in dark current calibrations (autodark; see Sect.~\ref{s:cameras}). Measurements were done at room temperature, with short breaks between differently exposed stacks to prevent overheating the sensor. However, sensor temperatures were not obtained from the camera software.

A separate data set consisting of 96 images taken with 4 seconds between each on the iPhone SE, during which the entire device palpably warmed up, was analyzed to identify thermal effects. Pearson $r$ correlations between response and time stamps (as a proxy for temperature) were calculated for the individual pixels. These $r$ values were well-described by a normal distribution with $\mu = 0.00$ and $\sigma = 0.10$, indicating that no strong relation exists between temperature and residual dark current. However, we note that again no direct sensor temperatures could be obtained.

In both cameras, a small residual (positive or negative) dark current signal was observed. Most pixels in both cameras had little dark current (RMS <2 ADU s$^{-1}$, 99.9th percentile of absolute values <6 ADU s$^{-1}$), though notable outliers were found, such as >300 pixels in our Galaxy S8 with dark current >50 ADU s$^{-1}$. The residual dark current decreased at higher ISO speeds, similar to RON and bias variations (Sect.~\ref{ss:results:bias_ron}), but showed no large-scale patterns.

These results show that autodarking accurately corrects most pixels, but is inadequate for outliers. Since autodarking is built into camera chips, it cannot be disabled. For outliers and in low-light conditions, it should be augmented with a manual dark current correction. As with bias variations, the dark current map $D(x,y,ISO)$ is used in Eq.~\eqref{e:radiometry} for low-light conditions, but an approximation is adequate for bright conditions. For autodarked cameras like the ones tested here, a mean value of $D = 0$ ADU s$^{-1}$ is assumed, and the RMS variation incorporated into the error budget. Outliers may be masked in either case.

\subsection{ISO speed} \label{ss:results:iso}

The normalization of data at different ISO speeds was measured using the methods from Sect.~\ref{ss:methods:iso} on two smartphones (iPhone SE and Galaxy S8) and one digicam (Nikon D5300).

\begin{figure}[ht]
	\centering
    \includegraphics[width=0.5\textwidth]{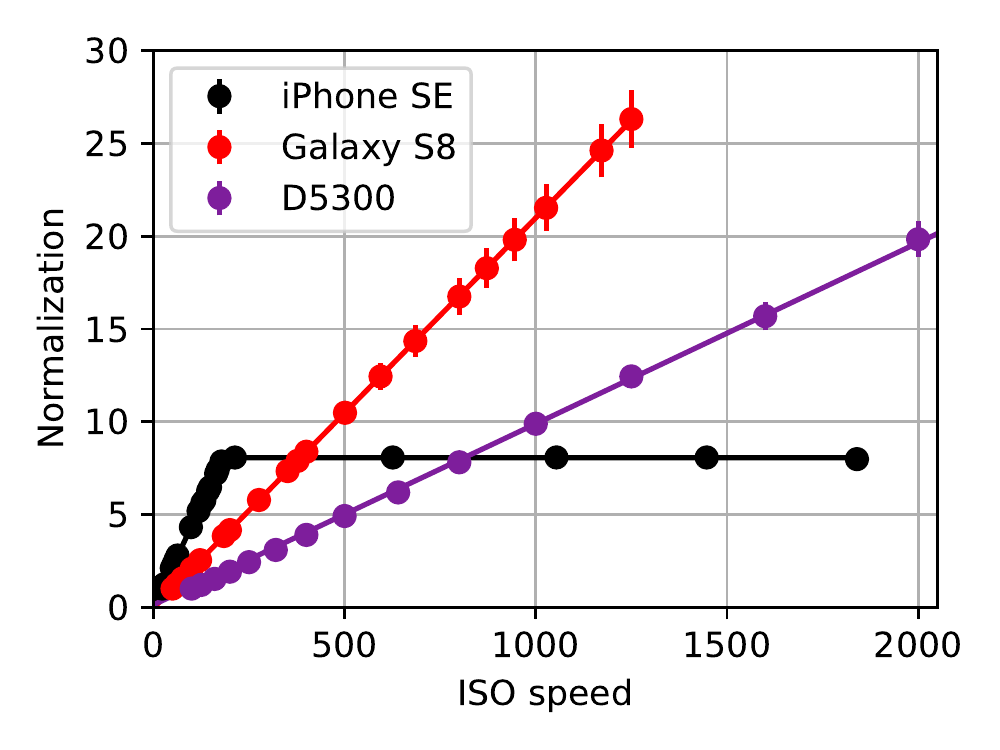}
    \caption{ISO speed normalization for the iPhone SE, Samsung Galaxy S8, and Nikon D5300. Dots indicate means of image stacks divided by the mean value per pixel at the lowest ISO speed. Lines indicate the best-fitting relationships.}
    \label{f:ISO}
\end{figure}

The measured and best-fit normalization curves are depicted in Fig.~\ref{f:ISO}. The Nikon D5300 and Galaxy S8 were best fit with a single linear relation, while the iPhone SE curve is clipped at ISO 184. This clipping is not due to image saturation, as none of the pixels in any image reached saturation. The linear part of the iPhone SE relation passes through the origin, while the Nikon D5300 and Galaxy S8 curves do not, instead showing significant (>5\%) systematic errors when using the simplest mathematical model (zero offset and slope 1/minimum ISO speed).

The clipping of the iPhone SE curve likely corresponds to a transition from purely analog to purely digital gain. However, data from the Camera2 API on the Galaxy S8 indicated that it too uses digital gain, at ISO speeds >640. This suggests that there are different implementations of gain for RAW photography.

The large observed differences in ISO speed normalization can lead to significant systematic errors when combining data taken at different ISO speeds, if not adequately calibrated. Data are normalized by dividing by $N$, as expressed in Eq.~\eqref{e:radiometry}.

\subsection{Gain} \label{ss:results:gain}
The methods from Sect.~\ref{ss:methods:gain} were used to characterize inter-pixel gain variations in two smartphones (iPhone SE and Galaxy S8).

\begin{figure}[ht]
	\centering
    \includegraphics[width=\textwidth]{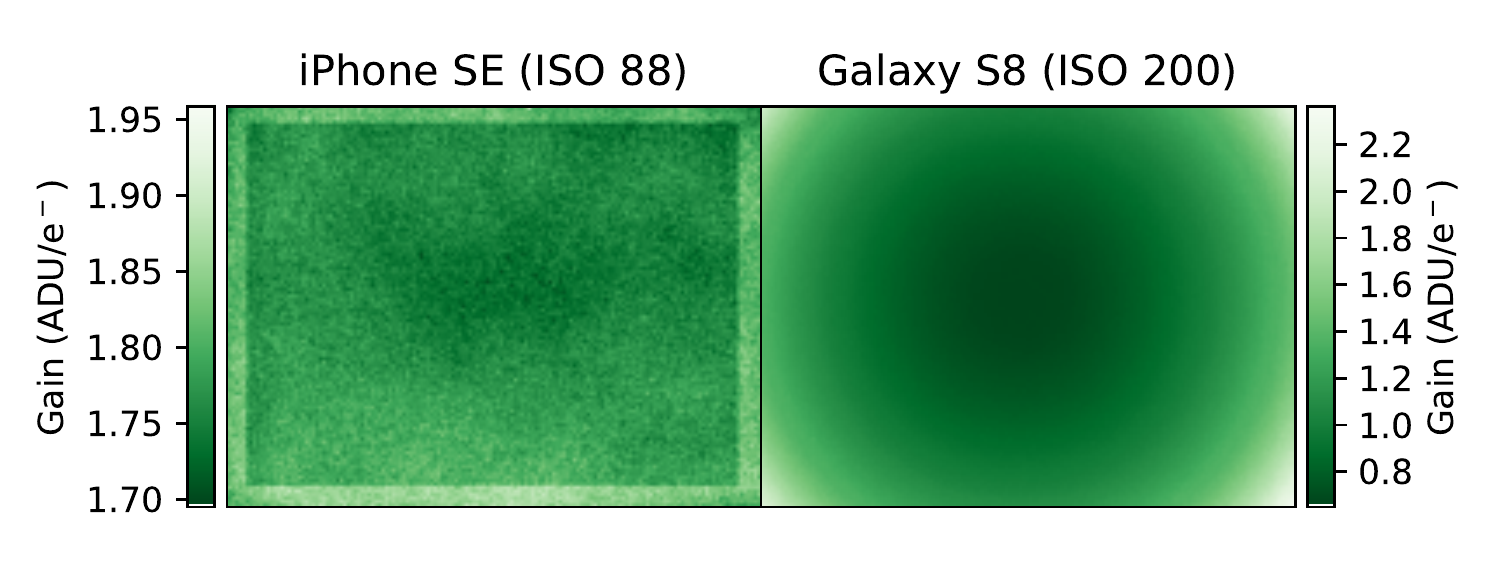}
    \caption{Gain values of G pixels in the iPhone SE (left; ISO speed 88) and Galaxy S8 (right; ISO speed 200) sensors. Darker colors indicate lower gain values. A two-dimensional Gaussian filter ($\sigma = 5$) has been applied to better visualize large-scale fluctuations. The iPhone SE patterns are similar to the read noise shown in Fig.~\ref{f:ron}.}
    \label{f:gain_map}
\end{figure}

Significant inter-pixel variations were observed, as shown in Fig.~\ref{f:gain_map} for the G pixels in both cameras. Since the measurement protocol is applied on the individual pixel level, the observed variations are only due to differences in gain, rather than external factors such as vignetting effects. The iPhone SE showed small variations, with higher gain values at the edges and lower values in the center. This pattern is similar to that seen in Fig.~\ref{f:ron}, suggesting a common origin. Conversely, on the Galaxy S8 a concentric pattern with a very wide range is clearly visible, likely intended as a first-order vignetting correction. Both showed similar ranges in gain (0.58--2.74 and 0.59--2.45 ADU/e$^-$, respectively), though on the iPhone SE most variations were on small scales and thus are not visible in the smoothed image.

\begin{figure}[ht]
	\centering
    \includegraphics[width=\textwidth]{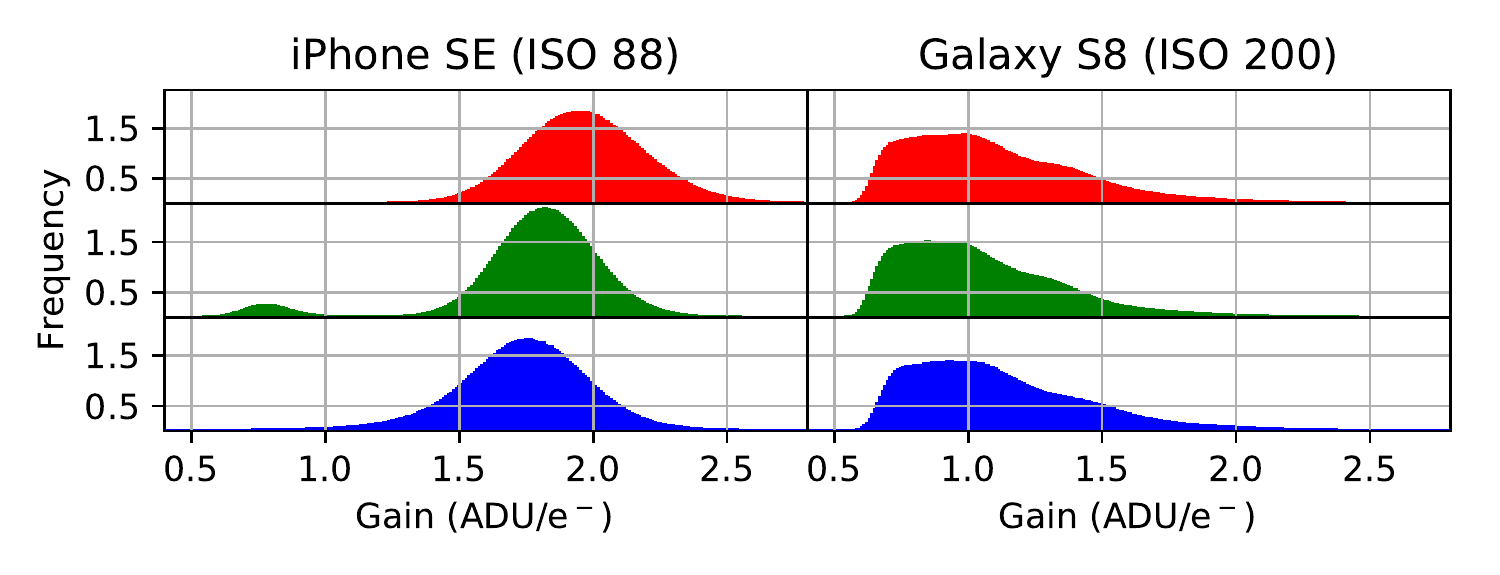}
    \caption{Histogram of gain values in the R (top), G and G$_2$ (middle), and B pixels (bottom) in the iPhone SE (left; ISO speed 88) and Galaxy S8 (right; ISO speed 200) sensors. The vertical axes were normalized to account for the different amounts of pixels.}
    \label{f:gain_hist}
\end{figure}

Histograms of gain values for both cameras are shown in Fig.~\ref{f:gain_hist}. Inter-filter differences are small in the Galaxy S8 but obvious in the iPhone SE. In particular, the R, G, and B distributions in the latter clearly have different mean values and widths (means and standard deviations of $1.97 \pm 0.24$, $1.78 \pm 0.29$, and $1.73 \pm 0.30$ ADU/e$^-$, respectively). Furthermore, the G distribution is bimodal while both others are unimodal; no significant differences between the $G$ and $G_2$ gain distributions were found, so this is not the cause for the observed bimodality.

The observed gain variations are significant and provide insights into camera design and fabrication, specifically the origins of some of the phenomena seen in flat-field data (Sect.~\ref{ss:results:flat}). However, they are not necessary in the radiometric calibration of data, since our flat-field correction method (Sects.~\ref{ss:methods:flat} and \ref{ss:results:flat}) accounts for these inter-pixel gain variations as well as all other differences in sensitivity, such as vignetting, as discussed in Sect.~\ref{ss:methods:calibration}.

\subsection{Flat-field correction} \label{ss:results:flat}

Three smartphones (iPhone SE and 6S, and Galaxy S8) and one UAV (DJI Phantom Pro 4) were flat-fielded using an integrating sphere as described in Sect.~\ref{ss:methods:flat}. An aperture of $f/2.8$ was used for the Phantom Pro 4, and on each device the maximum focus was used. 300 images were taken with the iPhone SE and Galaxy S8, 224 with the Phantom Pro 4, and 30 with the iPhone 6S. The latter was flat-fielded using a different set-up, with a Newport 819D-SF-4 integrating sphere and taking only 30 images as this was sufficient for an SNR >3 in >99\% of its pixels.

\begin{figure}[ht]
	\centering
    \includegraphics[width=\textwidth]{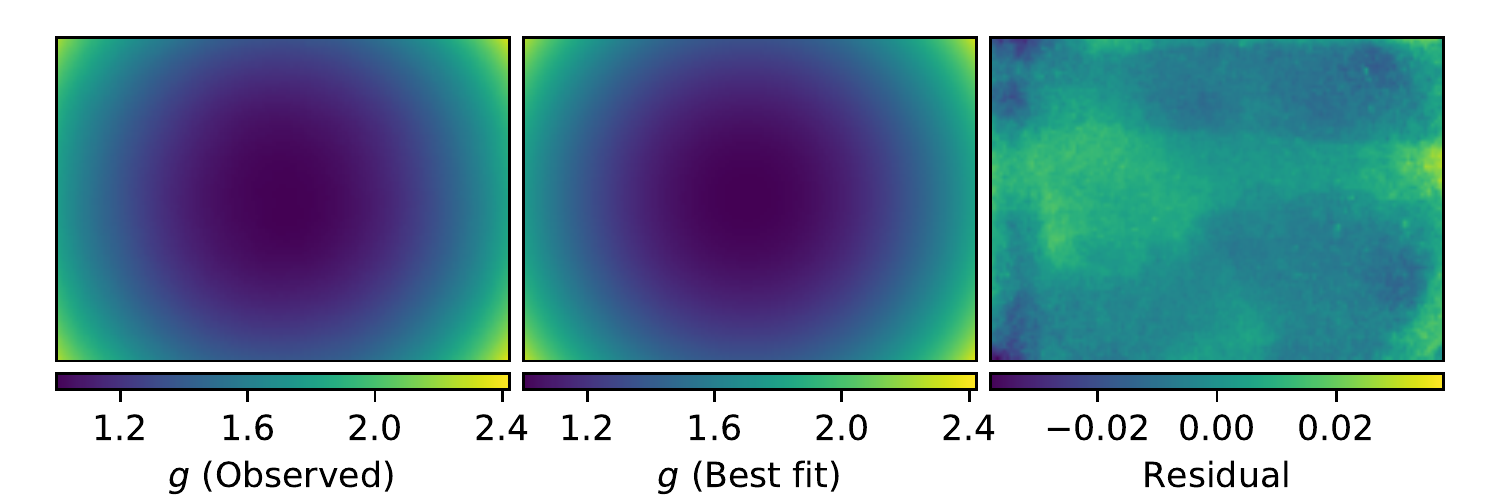}
    \caption{Flat-field correction factor $g$ for the iPhone SE camera. From left to right: observed values (inverse of observed relative sensitivity), best-fitting DNG model, and residuals.}
    \label{f:flat}
\end{figure}

Significant vignetting was found in all cameras. The observed correction factors of the iPhone SE, the best-fitting model, and residuals between the two are shown in Fig.~\ref{f:flat}. The smooth pattern suggests optical vignetting is the main flat-field component; the same is true in the iPhone 6S and Galaxy S8. The Phantom Pro 4 data showed an additional steep cut-off near the corners, suggesting mechanical vignetting. To counteract the latter, the outermost 250 pixels on all sides of the images from all cameras were removed prior to further analysis. Correction factors up to 2.42 (iPhone SE), 2.03 (iPhone 6S), 1.43 (Galaxy S8), and 2.79 (Phantom Pro 4) were observed. No significant chromatic differences were found, so the recombined data were used instead of separate RGBG$_2$ data.

As seen in Fig.~\ref{f:flat}, the DNG model fitted the data well with only small residuals remaining. The RMS of the residuals, normalized to the unsmoothed observed values, was 1.5\% (iPhone SE), 1.4\% (Galaxy S8), 3.1\% (iPhone 6S), and 2.0\% (Phantom Pro 4). These differences drop to $\leq$0.7\% on all cameras when using the spatially smoothed data, implying that they are mostly due to pixel-by-pixel variations and noise in the observations. These small residuals show that the DNG model is an adequate approximation for most applications; a pixel-by-pixel map per device is necessary only if sub-percent precision is required. Estimated errors in the model were <0.01 on the polynomial coefficients and <$10^{-5}$ on the optical center (in relative coordinates) for all cameras. Anomalous dots can be seen throughout the difference image in Fig.~\ref{f:flat}, possibly due to dust particles or inter-pixel gain variations (Sect.~\ref{ss:results:gain}). 

Since iOS also uses the DNG model for its internal vignetting correction, a direct comparison between correction models for the iPhone SE was made. The RMS relative residual between our smoothed data and the internal model was 5.9\%, more than 10 times that of our model (0.5\%). While the iOS model is symmetric ($\hat{c}_x = \hat{c}_y = 0.5$), ours had a slight offset ($\hat{c}_x = 0.494226(1)$ and $\hat{c}_y = 0.503718(2)$). The polynomial coefficients all differed by >400$\sigma$, with $\sigma$ the standard error on our model derived by the fitting routine. Finally, the RMS difference between the models per pixel was 5.7\%.


The three alternate methods described in Sect.~\ref{ss:methods:flat} were tested on the Galaxy S8. 40 images of the overcast sky were taken, as well as 40 of the sun and 50 of a monitor with a paper diffuser. The Galaxy S8 was used because its integrating sphere data show a large asymmetry ($\hat{c}_x = 0.449391(5), \hat{c}_y = 0.426436(9)$), providing a simple comparison metric. The RMS difference between the smoothed data from the integrating sphere and alternative methods relative to the sphere data were 4\%, 4\%, and 5\%, respectively. The best-fitting optical centers of all three data sets differed significantly both from the sphere data and from each other ($\hat{c}_x = 0.53447(1), 0.501989(4), 0.490794(4)$ and $\hat{c}_y = 0.38837(2), 0.449426(7), 0.477590(7)$, for the sky, sun, and monitor methods, respectively). This causes a typical systematic error on the order of 5\% in all three cases. Finally, six replicate measurement sets (50 images each) were taken using the monitor method to assess the effects of nonuniformities in the paper diffusers, generating a correction model for each set. The typical variation, expressed as the RMS of the standard deviation per pixel relative to the mean value per pixel, was 3\%, smaller than the typical deviations between the do-it-yourself methods and ground truth data. The effect of paper nonuniformities thus does not significantly impact the quality of do-it-yourself data.

The flat-field correction is incorporated in the radiometric correction expressed in Eq.~\eqref{e:radiometry} as the factor $g = g(x,y)$. For cameras with a fixed aperture, such as smartphones, one measurement is sufficient; otherwise, $g$ varies with aperture. This corrects for the systematic error induced by flat-fielding effects but pixels at the edges still receive fewer photons than those in the center. The former correspondingly have a smaller SNR due to shot noise, scaling as $SNR \propto g^{-1/2}$. Therefore, objects of interest are preferably imaged near the optical center of the camera.

\subsection{Spectral response} \label{ss:results:spectral}

Two smartphones (iPhone SE and Galaxy S8) and one UAV (DJI Phantom Pro 4) were spectrally calibrated using a monochromator, and the iPhone SE using iSPEX, as described in Sect.~\ref{ss:methods:spectral}.

\begin{figure}[ht]
	\centering
    \includegraphics[width=\textwidth]{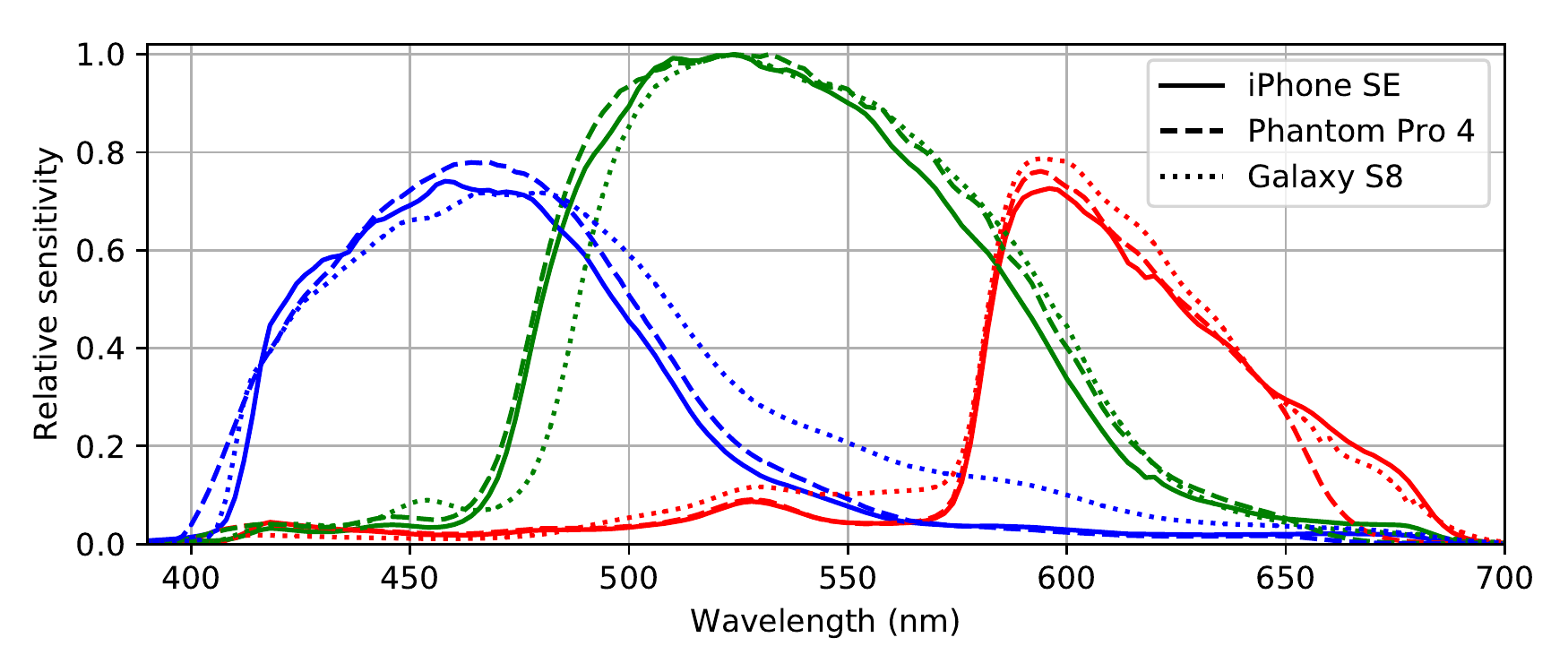}
    \caption{Spectral response curves of the iPhone SE, Galaxy S8, and Phantom Pro 4, derived from monochromator data. The responses are normalized to the global maximum per camera, giving relative sensitivities. G is the average of the G and G$_2$ responses over the wavelength axis, since no significant differences were found. RMS errors are $\leq$0.005.}
    \label{f:monochromator}
\end{figure}

Figure~\ref{f:monochromator} shows the normalized spectral response curves derived from the monochromator data, calibrated to the spectral throughput of the monochromator and spectral irradiance of the light source. This calibration was done by measuring its output under the same conditions as during the measurements, using a pre-calibrated silicon photodiode. Parts of the spectra were measured with different exposure settings and monochromator filters; these were first calibrated and then normalized and averaged on overlapping sections. The peak response wavelengths and effective bandwidths of the RGBG$_2$ filters in the different cameras are given in Table~\ref{t:spectral}.

\begin{table}[ht]
    \centering
    \begin{tabular}{l|rrrrrrrr}
         Camera        & $\lambda_{P, R}$ & $\Lambda_R$ & $\lambda_{P, G}$ & $\Lambda_G$ & $\lambda_{P, G2}$ & $\Lambda_{G2}$ & $\lambda_{P, B}$ & $\Lambda_B$  \\ \hline
         iPhone SE     & 596              & 72          & 524              & 110         & 524               & 109            & 458              & 93           \\
         Galaxy S8     & 594              & 73          & 524              & 109         & 524               & 109            & 468              & 117          \\
         Phantom Pro 4 & 594              & 65          & 524              & 115         & 532               & 116            & 468              & 94           \\
    \end{tabular}
    \caption{Peak response wavelength $\lambda_{P, \mathcal{C}}$ and effective spectral bandwidth $\Lambda_\mathcal{C}$ of each filter in the three cameras, derived from monochromator measurements. All values are in nm.}
    \label{t:spectral}
\end{table}

Some similarities and differences between the cameras are immediately obvious from Fig.~\ref{f:monochromator} and Table~\ref{t:spectral}. Notably, no significant differences between G and G$_2$ were found in any camera (RMS differences $\leq$0.003); the different peak wavelength for the Phantom Pro 4 is likely due to noise. The peak response wavelengths are very similar or even identical between cameras, as are the effective bandwidths, with two notable exceptions. The Galaxy S8 B filter is significantly broader than the others, with a comparatively high response at $\lambda > 500$ nm. Conversely, the Phantom Pro 4 has a relatively narrow R filters due to its NIR cut-off around 670 nm rather than 680 nm. Moreover, the R filters in all three cameras show a secondary peak around 535 nm and nearly identical responses between 570--650 nm.

\begin{figure}[ht]
	\centering
    \includegraphics[width=\textwidth]{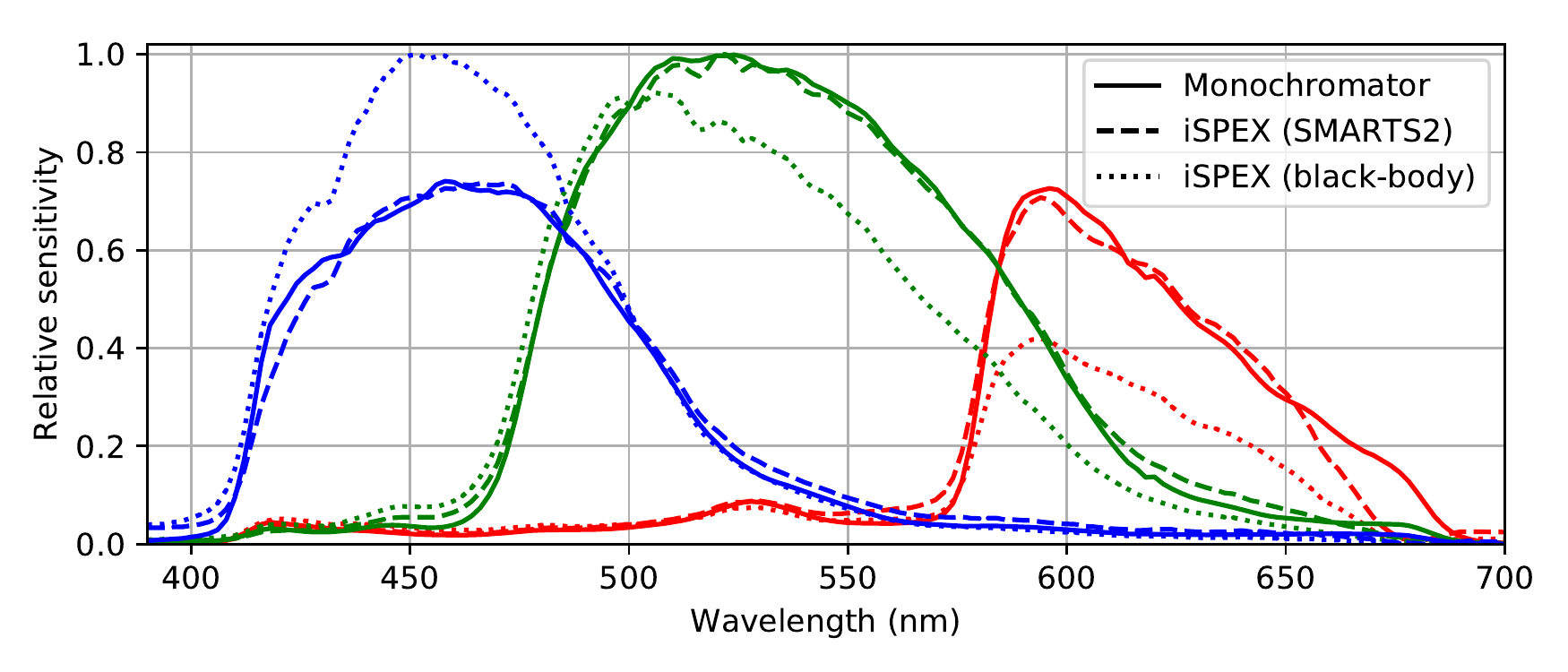}
    \caption{Comparison of the iPhone SE spectral response curves measured with the monochromator and iSPEX. iSPEX data are normalized using a 5777 K black-body and a SMARTS2 model, as described in Sect.~\ref{ss:methods:spectral}.}
    \label{f:ispex}
\end{figure}

The spectral response curves measured with iSPEX, shown in Fig.~\ref{f:ispex}, were similar to those derived from the monochromator data but showed small though significant systematic differences. No significant differences were found between narrow- and broad-slit spectra, so these were averaged. RMS differences between iSPEX- and monochromator-derived responses were 0.04, 0.02, and 0.02 (SMARTS2 normalization) and 0.12, 0.11, and 0.10 (black-body normalization), in RGB respectively. The black-body under-estimated the irradiance <500 nm and over-estimated it >500 nm compared to the SMARTS2 model, resulting in large deviations in the retrieved spectral response. The RMS difference between the monochromator-derived and black-body-normalized iSPEX-derived spectral responses could be reduced to 0.05, 0.11, and 0.04 by multiplying each filter with an empirical constant. However, systematic differences >0.2 remained in the G filter at wavelengths of 500-600 nm. Conversely, the SMARTS2-normalized iSPEX-derived spectral responses only showed a significant systematic difference compared to monochromator data at wavelengths >650 nm, the origins of which are unclear.

The observed differences between devices have important implications for RGB color measurements and spectroscopy, for example for color measurements as discussed in Sect.~\ref{s:intro}. The effective spectral bandwidths are incorporated into the radiometric calibration of imaging data as described in Sect.~\ref{ss:methods:calibration}. Furthermore, smartphone spectrometers naturally require calibration for the spectral response of the camera, as expressed in Eq.~\eqref{e:radiometry_spectral}.

\subsection{SPECTACLE database} \label{ss:results:database}

To facilitate the use of consumer cameras in scientific projects and improve future compatibility, we have created the SPECTACLE (Standardised Photographic Equipment Calibration Technique And CataLoguE) database. It includes the calibration data required for radiometric corrections (Sect.~\ref{ss:methods:calibration}), for the most popular cameras. The data are given in standardized formats, split into three categories (device, camera, and software) to minimize the amount of data required. For example, two devices using the same camera module have the same spectral response curves and flat-field response, while software parameters such as bias and ISO speed settings vary. The former can thus be combined while keeping the latter separate. Since the properties of a camera may change with firmware updates or changes in manufacturing, database entries may be split according to device version, rather than assuming devices of the same model are clones. Finally, given calibration data for multiple identical devices, statistics on variations within a camera model may be included. The open design of the SPECTACLE database, based on the Parse platform, allows anyone to use or contribute data, particularly using the calibration apps we have developed. Submitted data are currently curated by the authors to ensure their quality. As the database grows, community curation or automated curation based on outlier analysis may become preferable. SPECTACLE can be accessed at \url{http://spectacle.ddq.nl/}.

\section{Discussion \& conclusions} \label{s:discussion}

In this work, we have presented a standardized calibration methodology for the most important factors limiting the quality of consumer camera data, the first to our knowledge. Furthermore, we have developed the SPECTACLE database, containing calibration data for the most popular devices. The standardized methodology and the SPECTACLE database have the potential to improve the sustainability of projects using these cameras, by simplifying their calibration and the use of multiple camera models.

The main difference between our approach and those in much of the literature is the use of RAW data. Software constraints previously forced the use of JPEG data, which are compressed and heavily processed, introducing systematic effects that negatively affect the data quality and are difficult to calibrate \cite{Leeuw2018hydrocolor, Igoe2018ozone, Rateni2017smartphonefoodreview, Zhang2016GFresnel, Novoa2015WACODI, Snik2014ispex, Skandarajah2014microscope, Bongiorno2013spectralresponse, Kreuter2009allskypolarimetry}. The desire to use RAW data has been expressed widely in the literature \cite{Leeuw2018hydrocolor, Rateni2017smartphonefoodreview, Pagnutti2017PiCamera, McCracken2016resourcelimited, Novoa2015WACODI, Nguyen2014raw2raw, Bongiorno2013spectralresponse, Lebourgeois2008commercialcameras}, and their superiority is clearly demonstrated by the highly linear response and larger dynamic range found in Sect.~\ref{ss:results:linearity}. The former is especially notable since nonlinearity and the associated gamma correction are among the most cited problems of JPEG data \cite{SanchezdeMiguel2019artificiallight, Coburn2018remotesensingcameras, Leeuw2018hydrocolor, Ding2018mHealth, Xu2018communication, Wang2017herbicide, Novoa2015WACODI, Sumriddetchkajorn2014chlorine, Snik2014ispex, Skandarajah2014microscope, Bongiorno2013spectralresponse, Charriere2013rgbmicroscope, Kreuter2010skypolarization, Kreuter2009allskypolarimetry, Lebourgeois2008commercialcameras, Cescatti2007canopycamera}. While JPEG nonlinearity corrections exist, either fully empirical or based on the sRGB standard \cite{Coburn2018remotesensingcameras, Novoa2015WACODI, Skandarajah2014microscope}, the wide (1.7--2.6) variations in gamma and large (>30\%) deviations from sRGB profiles shown in Sect.~\ref{ss:results:linearity} and Fig.~\ref{f:jpeg} indicate that these are inaccurate and difficult to generalize. The highly linear nature of RAW data was previously demonstrated in \cite{Pagnutti2017PiCamera, Manakov2016evaluation, Darrodi2015spectralsensitivity} and may be a result of internal linearity corrections in the CMOS chip \cite{Zhang2011linearCMOS}. Furthermore, RAW data are not affected by white balance, a color correction in JPEG processing which severely affects colorimetric measurements, is difficult to calibrate, and differs strongly between measurements and cameras \cite{SanchezdeMiguel2019artificiallight, Gallagher2018hydrocolor, Friedrichs2017SmartFluo, Rateni2017smartphonefoodreview, McCracken2016resourcelimited, Novoa2015WACODI, Snik2014ispex, Skandarajah2014microscope, Goddijn2009fundamentals, Kim2008radiometryvignetting}. This variable gamma correction and white balance make it impossible to invert the JPEG algorithm and recover RAW data. However, RAW data are no panacea, since they still require further calibrations. Furthermore, not all consumer cameras support RAW imagery, especially low-end smartphones; hence the low adoption rate in literature until now. Still, we consider the linearity, larger dynamic range, and lack of unknown post-processing affecting the data worth relying on RAW data, especially in a market trending towards broader support.

Inter-pixel and inter-device bias variations and read noise were found to be small in general ($\sigma$<1 for bias variations, mean RON <3 ADU), though with large outliers (Sect.~\ref{ss:results:bias_ron}). These distributions are similar to those found in several smartphones \cite{Whiteson2016cosmicrays} and a Raspberry Pi camera \cite{Pagnutti2017PiCamera}, though neither work distinguishes between bias variations, read noise, and dark current. The large-scale patterns seen in Fig.~\ref{f:ron} were not found in the literature. Their cause is unclear, though correlations with inter-pixel gain variations (Sect.~\ref{ss:results:gain}) suggest a common origin. Ultimately, since both phenomena are small, for most applications these patterns are merely a curiosity and an estimate in the error budget and masking of outliers is sufficient for further radiometric calibrations (Sect.~\ref{ss:methods:calibration}).

While dark current has been implicated in the literature as a major noise source \cite{Igoe2018ozone, Kim2017smartphonebioluminescence, Turner2017UVB, Pagnutti2017PiCamera, Whiteson2016cosmicrays, Berra2015UAVs, Igoe2014dark}, the results presented in Sect.~\ref{ss:results:dark} indicate that it is actually typically quite minor. The RMS dark current in the iPhone SE and Galaxy S8 (<2 ADU s$^{-1}$) is similar to values found in \cite{Coburn2018remotesensingcameras, Igoe2018ozone, Turner2017UVB, Igoe2014dark}, while we found larger outliers, such as >300 pixels with >50 ADU s$^{-1}$ in our Galaxy S8. Similarly to \cite{Igoe2014dark}, no significant relationship was found between temperature and residual dark current, though this experiment should be repeated under more controlled conditions and using internal sensor temperatures to draw strong conclusions. In general, a quantitative comparison with the literature is difficult, since those studies used JPEG data, not RAW. While our sample of two cameras is insufficient to draw broad conclusions, these results suggest that dark current is less important than previously thought. As discussed in Sect.~\ref{ss:results:dark} and similarly to the aforementioned bias and RON variations, extensive characterization of the dark current in individual pixels is necessary for low-light applications and spectroscopy as these are significantly affected by a few `bad' pixels. Conversely, for bright-light conditions the dark response is typically negligible and an ensemble estimate in the error budget and masking of outliers are sufficient.

ISO speed normalization is typically done by simply dividing digital values by the ISO speed \cite{Leeuw2018hydrocolor, Zhao2009spectralsensitivity}, but the results presented in Sect.~\ref{ss:results:iso} and Fig.~\ref{f:ISO} contradict the validity of this method. This discrepancy was also identified in \cite{Sigernes2009absolutesensitivity}. Observed relations differ significantly from the na\"ive linear model in shape, offset and slope. For example, differences between the two models of >5\% were found in the Galaxy S8. More extremely, the expected and observed normalization factor at ISO speed 1840 on the iPhone SE differ by a factor of 10. Moreover, Android documentation suggests that more complex curves with mixed analog and digital gain may also be in use. Thus, to prevent similar systematic errors, either a single ISO speed per device must be used or these relations must be calibrated.

Significant inter-pixel gain variations were found in Sect.~\ref{ss:results:gain}, as shown in Figs.~\ref{f:gain_map} and \ref{f:gain_hist}. The Galaxy S8 showed a strong radial pattern, likely intended as a first-order vignetting correction; this was not seen in the iPhone SE. Conversely, gain values in the latter differed significantly between color filters. This may be a color correction called analog white balance, which is described in the DNG standard \cite{DNG1.4.0.0}; however, in this case it is not clear why significant inter-pixel variations exist. No previous discussion of such variations in gain in a consumer camera was found in the literature. Typically, an equal gain in all pixels is assumed in absolute radiometric calibrations \cite{Pagnutti2017PiCamera, Sigernes2009absolutesensitivity} but the variations found here cast doubt on the generality of this assumption.

Strong flat-field effects were found in Sect.~\ref{ss:results:flat}, with correction factors up to 2.79. Similarly large correction factors have been found for other cameras, for instance approximately 2.8 in a Canon EOS 400D \cite{Lebourgeois2008commercialcameras} and 650D \cite{Kordecki2017vignettingSPIE}, 4 in a Raspberry Pi camera \cite{Pagnutti2017PiCamera}, 1.8 in a Canon EOS 10D \cite{Goldman2010vignetting}, and 1.5 in a Nikon E775 \cite{Zheng2009vignetting}. It should be noted that vignetting is highly aperture-dependent and thus these correction factors will change with varying apertures \cite{Goldman2010vignetting}. Interestingly, we did not find the large chromatic differences described in \cite{Pagnutti2017PiCamera, Lebourgeois2008commercialcameras}. Notably, the Galaxy S8 showed a much weaker vignetting effect ($g_{max} = 1.43$) than the other cameras ($g_{max} > 2$), likely due to the aforementioned inter-pixel gain variations. These may also explain the strong asymmetry ($\hat{c}_x = 0.449391(5), \hat{c}_y = 0.426436(9)$) seen in the Galaxy S8, due to the main symmetrical component having been corrected already. 

The 7-parameter vignetting model described in the DNG standard \cite{DNG1.4.0.0} fits our data very well (RMSE $\leq$3.1\% for raw data, $\leq$0.7\% for smoothed data), without significant systematic differences. Since the typical difference between observed and modeled corrections is small, pixel-by-pixel flat-fielding is necessary only for applications requiring sub-percent precision. For those, a flat-field map would be made for each individual device, rather than using the same map for multiple devices of the same model. Flat-field measurements of the latter could be used to quantify typical variations in flat-field response among identical devices and further determine when pixel-by-pixel or modeled flat-field corrections are preferable. The DNG model is also used for internal vignetting correction in iOS. While this correction is sometimes considered a major advantage of JPEG data over RAW data, the internal model of the iPhone SE was shown to be significantly less accurate (RMSE $ = 5.9$\%) than one based on our data (RMSE $ = 0.5$\%), contradicting this notion. Moreover, residual vignetting effects up to 15\% have been observed in JPEG data \cite{Coburn2018remotesensingcameras}. A comparison to the internal correction data in Android smartphones, consisting of pixel-by-pixel look-up tables, has not yet been done since these data are relatively difficult to access.

Finally, three simpler alternative flat-fielding methods were tested, namely imaging the sky, the sun, and a computer monitor, as described in Sect.~\ref{ss:methods:flat}. Applied on the Galaxy S8, data from these methods differed from the integrating sphere data by $\leq$5\% RMS. These errors mainly result from a difference in the location of the optical center. The cause of these discrepancies is unclear, though insufficiently isotropic light sources are an obvious explanation. Nevertheless, the RMS difference of $\leq$5\% is small compared to the overall flat-field correction of up to 179\% and better than the internal correction of the iPhone (RMS 5.9\%). These methods thus serve as a useful first estimate for the flat-field correction in the absence of integrating sphere data. As discussed in Sect.~\ref{s:cameras}, many further alternative flat-fielding methods exist \cite{Kordecki2017vignettingSPIE, Silva2016vignetting, Goldman2010vignetting, Zheng2009vignetting, Kim2008radiometryvignetting, Lebourgeois2008commercialcameras, Yu2004vignetting}. Our data may be useful as a ground truth for a thorough comparison of such methods akin to \cite{Kordecki2017vignettingSPIE, Manakov2016evaluation}.

The spectral responses found in Sect.~\ref{ss:results:spectral} and shown in Fig.~\ref{f:monochromator} agree well with those found in the literature \cite{SanchezdeMiguel2019artificiallight, Coburn2018remotesensingcameras, Leeuw2018hydrocolor, Pagnutti2017PiCamera, Manakov2016evaluation, Berra2015UAVs, Darrodi2015spectralsensitivity, Javorsek2015nikonspectra, Jiang2013spectralsensitivity, Bongiorno2013spectralresponse, Zhao2009spectralsensitivity, Sigernes2009absolutesensitivity, Goddijn2009fundamentals, Lebourgeois2008commercialcameras}, with the RGB curves centered around 600, 520, and 470 nm, respectively. Notably, the strong secondary peaks seen in \cite{Leeuw2018hydrocolor, Coburn2018remotesensingcameras} were not found in our data and may be JPEG artefacts. Differences are mainly found in the wings, such as the NIR cut-offs \cite{Berra2015UAVs, Lebourgeois2008commercialcameras} and harmonics. The comparatively high response of the Galaxy S8 B filter at wavelengths >500 nm is also seen in the Nokia N900 \cite{Jiang2013spectralsensitivity} and Sony A7SII \cite{SanchezdeMiguel2019artificiallight}, and to a lesser extent the Galaxy S5 \cite{Leeuw2018hydrocolor}, but is otherwise uncommon. The early NIR cut-off of the Phantom Pro 4 appears to be similarly uncommon but not unique \cite{SanchezdeMiguel2019artificiallight, Leeuw2018hydrocolor, Berra2015UAVs, Jiang2013spectralsensitivity}. These differences again show the importance of spectral characterization for normalizing smartphone spectrometer data. Furthermore, the significant variations show that the common assumption of sRGB responses \cite{Novoa2015WACODI, Charriere2013rgbmicroscope} does not hold, as has been suggested previously \cite{Nguyen2014raw2raw}, and characterization of the spectral response is necessary to convert observed colors to color spaces such as CIE 1931 XYZ or CIELAB \cite{Jiang2013spectralsensitivity, Cheung2005multispectralcharacterization}. However, color measurements still depend on the incident light spectrum \cite{Cheung2005multispectralcharacterization}; hyperspectral measurements, for example with iSPEX \cite{Snik2014ispex}, and characterization of common light sources \cite{SanchezdeMiguel2019artificiallight, Tapia2018LICALamps} may provide valuable additional information. Finally, while no significant response was found at wavelengths <390 or >700 nm on our test cameras, it may be worthwhile in the future and the SPECTACLE database to use a spectral range of 380-780 nm to follow colorimetric standards \cite{Xu2018communication, Turner2017UVB, Cheung2005multispectralcharacterization}.

Spectral response measurements done with the iSPEX smartphone spectrometer \cite{Snik2014ispex} agreed well (RMS differences $\leq$0.04) with the monochromator measurements (Sect.~\ref{ss:results:spectral} and Fig.~\ref{f:ispex}). The only systematic difference was an under-estimation at wavelengths >650 nm, though it is unclear what causes this. The good agreement shows that iSPEX measurements are an adequate replacement for monochromator data if the latter are not available. This will be especially useful with the new iSPEX we are developing, which will also feature universal smartphone hardware interface. One downside of this method is that it requires an accurate solar reference spectrum. We used one generated with SMARTS2 \cite{Gueymard2001SMARTS2, Gueymard1995SMARTS2}; this model matches observed solar spectra very well but it is not very portable or user-friendly for non-expert users. A 5777 K black-body approximation was also used but reproduced the SMARTS2 spectrum poorly (RMSE of 34\%) and accurate spectral response curves could not be retrieved this way. A more portable model or set of standard spectra could improve the user-friendliness of this calibration method.

Further alternative methods for spectral response characterization include those based on multispectral measurements using computational methods to enhance their resolution \cite{Chaji2018spectralsensitivity, Finlayson2016spectralsensitivity, Yang2015LEDspectralresponse, Cheung2005multispectralcharacterization} or those using a linear variable edge filter \cite{Bongiorno2013spectralresponse}. However, the former are not sufficiently accurate \cite{Darrodi2015spectralsensitivity} while the latter is not necessarily more accessible than a monochromator. Our data may be used as a ground-truth for testing other methods akin to \cite{Darrodi2015spectralsensitivity} but with the advantage of smartphones being more accessible than the cameras used therein.

Finally, we have created the SPECTACLE database containing the calibration data described above. The aim of this database is to facilitate the use of consumer cameras in scientific projects by reducing the labor required for calibration. Data sets containing spectral responses \cite{Darrodi2015spectralsensitivity, Jiang2013spectralsensitivity} and extensive calibrations of single cameras \cite{Pagnutti2017PiCamera} have been published before but to our knowledge SPECTACLE is the first comprehensive, centralized spectral and radiometric calibration database. It is designed with community participation in mind, relying on volunteer contributions to become and remain complete in the rapidly evolving camera market. This will require a critical mass of users to maintain it, which is easier if more accessible calibration methods, like those discussed previously, can be used. We have kick-started this process with the calibrations done in this paper and will continue this while developing iSPEX.

Though extensive, our calibration methodology is not complete. The two most prominent missing components are geometric distortions and absolute radiometric calibration. The former are a well-known phenomenon with a large impact on image quality but relatively simple to measure and correct \cite{Tauro2016surfaceflow, Rasmussen2016UAVvegetation, Flynn2014UAVaquaticvegetation, Sumriddetchkajorn2014chlorine, Liang2014treemapping, Yu2004vignetting}. A parametric model for distortion is given in the DNG standard \cite{DNG1.4.0.0} and a comparison between measured distortions and the internal correction models of different cameras, similar to that done in Sect.~\ref{ss:results:flat} for vignetting corrections, may be used to determine the accuracy of the latter. Absolute radiometric calibration is extremely valuable for quantitative measurements, as described in Sect.~\ref{ss:methods:calibration}. In principle, our methods and calibration data contain most of the information required for this, bar a constant $K$. Absolute radiometric calibration of consumer cameras has been demonstrated before, notably in the Raspberry Pi camera \cite{Pagnutti2017PiCamera}, and Nikon D300 and Canon 40D \cite{Sigernes2009absolutesensitivity}, though only for a small number of devices. Another notable example is the Hukseflux Pyranometer app (Sect.~\ref{ss:methods:flat}) for measurements of solar irradiance, though it is intended for education and entertainment rather than scientific measurements. Finally, most of our calibrations were done on a single device, and differences between devices may exist, as shown in Fig.~\ref{f:ron}. Calibration of multiple devices per camera model would allow the characterization of these differences and the associated errors when using multiple devices. Additionally, differences may be introduced by changes in manufacturing or camera software. Characterization of different generations of the same model camera will be necessary to characterize these, which may result in separate entries in the SPECTACLE database for each camera version being necessary. However, the modular design of the SPECTACLE database makes it simple to extend. The simple, standardized calibration methods described in this work and the SPECTACLE database have the potential to greatly improve the data quality and sustainability of future scientific projects using consumer cameras.

\section*{Funding}
European Commission Horizon 2020 program (grant nr. 776480, MONOCLE and grant nr. 824603, ACTION)

\section*{Acknowledgments}

The authors wish to thank Molly and Chris MacLellan of the NERC Field Spectroscopy Facility for experimental help and invaluable insights in the flat-field and spectral response measurements. Figure~\ref{f:setup_linearity_polarizers} was drawn using \textit{draw.io}. Data analysis and visualization were done using the AstroPy, ExifRead, Matplotlib, NumPy, RawPy, and SciPy libraries for Python. Finally, the authors wish to thank the two anonymous reviewers for their thorough and constructive reviews. This project has received funding from the European Union's Horizon 2020 research and innovation programme under grant agreement No 776480.

\end{document}